\documentclass[
 reprint,twocolumn,
 groupedaddress,superscriptaddress,
 amsmath,amssymb,aps,prl,
]{revtex4-2}
\usepackage[shortlabels]{enumitem}
\usepackage{graphicx}
\usepackage{dcolumn}
\usepackage{bm}
\usepackage{hhline}
\usepackage{bbold}
\usepackage[shortlabels]{enumitem}

\begin{document}

\preprint{APS/123-QED}


\title{Exploiting disorder to probe spin and energy hydrodynamics}

\author{
Pai Peng,$^{1,*,\dag}$
Bingtian Ye,$^{2,3,*}$
Norman Y. Yao,$^{2,3}$ 
Paola Cappellaro,$^{4,5,\ddag}$
\\
\normalsize{$^{1}$Department of Electrical Engineering and Computer Science, Massachusetts Institute of Technology, Cambridge, MA 02139}\\
\normalsize{$^{2}$Department of Physics, University of California, Berkeley, CA 94720, USA}\\
\normalsize{$^{3}$Department of Physics, Harvard University, Cambridge, MA 02138, USA}\\
\normalsize{$^{4}$Department of Nuclear Science and Engineering, Massachusetts Institute of Technology, Cambridge, MA 02139, USA}\\
\normalsize{$^{5}$Research Laboratory of Electronics, Massachusetts Institute of Technology, Cambridge, MA 02139, USA}\\
\normalsize{$^*$These authors contributed equally to this work.}\\
\normalsize{$^\dag$To whom correspondence should be addressed; E-mail:  paipeng@mit.edu}\\
\normalsize{$^\ddag$To whom correspondence should be addressed; E-mail:  pcappell@mit.edu}\\
}

\date{\today}




\maketitle
\textbf{
An outstanding challenge in large-scale quantum platforms is to simultaneously achieve strong interactions, giving rise to the most interesting behaviors, and local addressing  ---that can probe them. 
In the context of correlated phases, local addressing enables one to directly probe the nature of the system's order.
Meanwhile, for out-of-equilibrium dynamics, such addressing allows the study of quantum information spreading and operator growth. 
%
%
Here, we introduce a novel technique that enables the measurement of local correlation functions, down to single-site resolution, despite access to only global controls. 
Our approach leverages the intrinsic disorder present in a solid-state spin ensemble to dephase the non-local components of the correlation function. 
Utilizing this toolset, we measure both the spin and energy transport in nuclear spin chains.
By tuning the interaction Hamiltonian via Floquet engineering, we investigate the cross-over between ballistic and diffusive hydrodynamics. 
Interestingly, when the system is both interacting and (nearly-)integrable, we observe the coexistence of diffusive spin transport with ballistic energy transport.
}

The complex dynamics of isolated quantum many-body systems are often amenable to a simple yet powerful description given by classical hydrodynamics \cite{halliwell1999decoherent,wyatt2005quantum,hartle2011quasiclassical,spohn2012large,birkhoff2015hydrodynamics,de2018hydrodynamic}. 
However, characterizing the nature of these hydrodynamical descriptions \cite{andreev2011hydrodynamic,vznidarivc2016diffusive,bertini2016transport,leviatan2017quantum,ye2020emergent,ljubotina2019kardar,ye2022universal,sommer2011universal,moll2016evidence,cepellotti2015phonon,crossno2016observation} and how they emerge from microscopic quantum dynamics remains an area of active pursuit \cite{agarwal2015anomalous,castro2016emergent,bertini2021finite,ilievski2017microscopic,gopalakrishnan2019kinetic,de2019diffusion,ilievski2021superuniversality,de2021stability,friedman2020diffusive}. 
Recently, this pursuit has seen tremendous advances owing to the development of large-scale quantum simulation platforms ranging from ultracold atoms and superconducting circuits, to solid-state spin systems~\cite{schemmer2019generalized,zu2021emergent,malvania2021generalized,wei2022quantum,joshi2022observing}. %

In order to control and probe many-body dynamics in such systems, one typically requires a combination of strong interactions and local manipulation. 
In the majority of platforms, these two features are in tension: Strong interactions arise when the constituent degrees of freedom are closely spaced, which in turn challenges the ability to perform
local measurements~\cite{Altman21, Bakr09}. 
The tension is particularly acute in solid-state platforms where electronic and nuclear spins can exhibit strong interactions only when spaced at nanometer length-scales. 
Here, we demonstrate that disorder, often times unavoidable in solids and long-considered detrimental for quantum coherence and transport, can be a powerful source of local control.
%
%
First, by dephasing a homogenous state using the disorder, we demonstrate the preparation of  states whose polarization on different sites is uncorrelated. %
%
Second, we show that single-site, spin-spin correlation functions can be directly measured using spin echo. 
%
The intuition behind our approach is the following -- owing to the lack of spatial correlations, non-local components of the correlation function are averaged out, leaving only a sum of autocorrelations. 
Applying our technique in the context of nuclear magnetic resonance, we demonstrate the direct observation and characterization of nanoscale spin and energy transport, without the need for magnetic field gradients, sub-diffraction techniques, or multiple spin species~\cite{zhang1998first,rittweger2009sted,maurer2010far,chen2013wide,pfender2014single,arai2015fourier,zu2021emergent}. 

\begin{figure*}[!htp]
    \centering
    \includegraphics[width=0.65\textwidth]{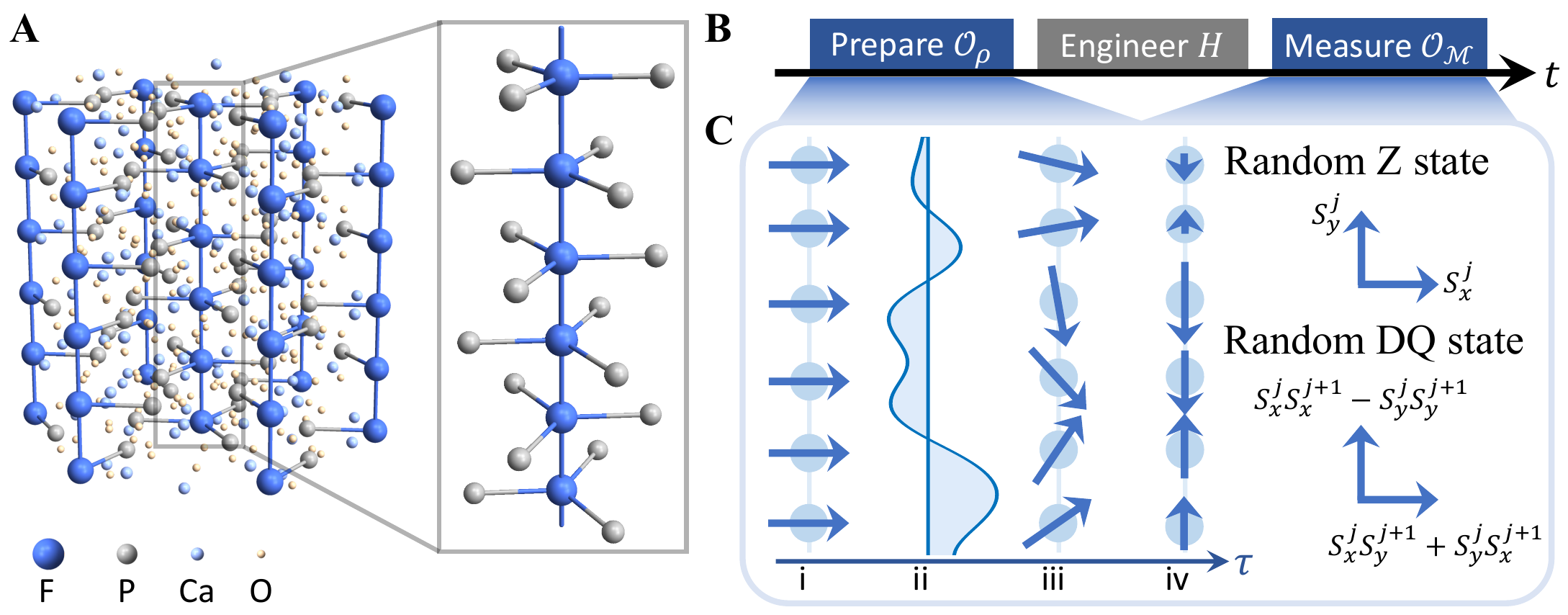}
    \caption{\textbf{Measuring local autocorrelations by utilizing global control and intrinsic on-site disorder.}
    (\textbf{A}) Chemical structure of Fluorapatite. $^{19}$F nuclear spins (blue balls) form a quasi-1D structure and can exhibit different classes of hydrodynamics under various effective Hamiltonians realized by Floquet engineering. $^{31}$P atoms (gray) provide intrinsic on-site disordered fields on $^{19}$F spins, which enables the preparation of random states and observables.
    (\textbf{B}) The experimental protocol to obtain local autocorrelations consists of three main stages. The central ingredient is to realize spatially uncorrelated random states and observables in the preparation and the measurement stages. Hamiltonian engineering enables varying the quantum dynamics.
    (\textbf{C}) The sequence to prepare random states and observables contains four steps: (i) initialize the system to a spatially homogeneous state with polarization along x axis, (ii) apply disordered field along z axis to encode (iii) local information into the spin phases, (iv) perform phase cycling to eliminate the residual homogeneous part. The  arrows represent spin operators whose bases are specified on the right for the random Zeeman state and the random DQ state, respectively. 
    To effectively measure spatially random observables, we apply the same sequence in reverse order to the final state before  measuring  the homogeneous magnetization.}
    \label{fig1}
\end{figure*}


\begin{figure*}[!htp]
    \centering
    \includegraphics[width=0.95\textwidth]{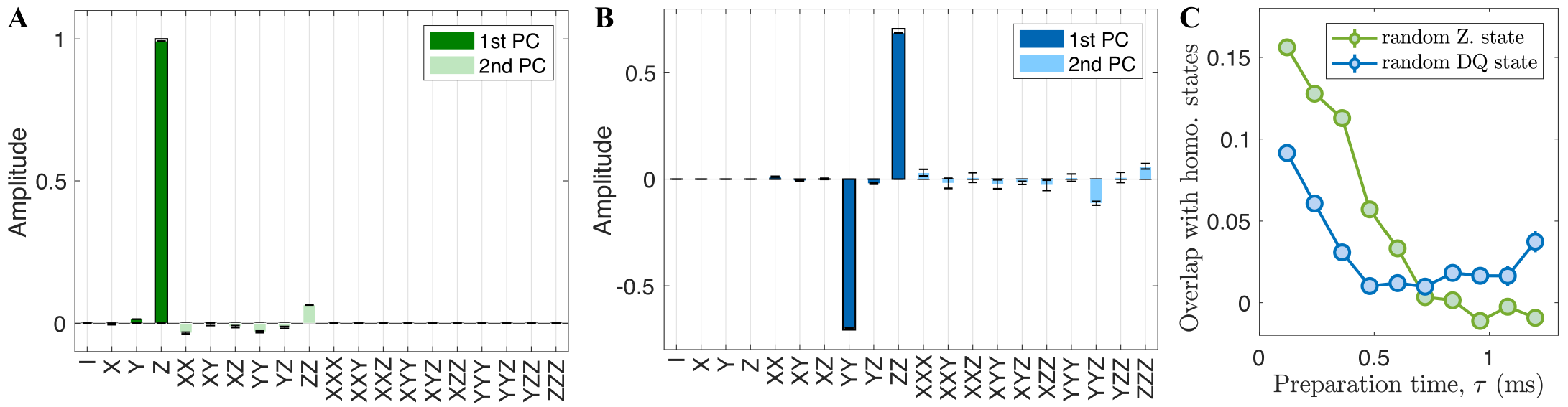}
    \caption{\textbf{Experimental verification of random initial state preparation}
    First two principal components (PC) of the experimentally prepared random Zeeman (\textbf{A}) and DQ states (\textbf{B}). The horizontal axis label stands for sum of all permutations of the corresponding spin operators, e.g. $XY$ corresponds to $\sqrt{2}(S_x^jS_y^{j+1}+S_y^jS_x^{j+1})$ normalized such that the Frobenius norm is $2^L$. The amplitude is weighted by the square root of the eigenvalue $\sqrt{\lambda_i}$. For random Zeeman state, $\lambda_1=0.985(1)$, $\lambda_2=0.0066(1)$; For random DQ state, $\lambda_1=0.963(9)$, $\lambda_2=0.0019(5)$. The eigenvalues are normalized such that $\sum_\mu \lambda_\mu=1$. The preparation time is 1.08~ms for random Zeeman state and 0.96~ms for random DQ state. The green and blue bars show the experimental results, the black wireframes show the ideal states. (\textbf{C}) Overlap of experimentally prepared random Zeeman (green) and DQ (blue) state with the corresponding homogeneous state quickly decays to zero. The overlap of two observables $\mathcal{O}_1,\mathcal{O}_2$ are defined as $\mathrm{Tr}(\mathcal{O}_1\mathcal{O}_2)/\sqrt{\mathrm{Tr}(\mathcal{O}_1\mathcal{O}_1)\mathrm{Tr}(\mathcal{O}_2\mathcal{O}_2)}$.
    }
    \label{fig2}
\end{figure*}

Our experiments are performed on $S=1/2$ $^{19}$F nuclear spins within a single crystal of fluorapatite.
The nuclear spins effectively form quasi-one-dimensional chains, since the inter-chain couplings are $\sim40$ times weaker than the intra-chain couplings [Fig.~\ref{fig1}A].
%
%
We place our sample in a 7~Tesla magnetic field along the [001] axis, which leads to a strong Z splitting that reduces the dipolar interaction between $^{19}$F spins to its secular form,
\begin{equation}
H_{\textrm{FF}}=\sum_{j<k}\frac{J}{2r^3_{jk}}(2S_z^jS_z^k-S_x^jS_x^k-S_y^jS_y^k),
\label{eq:F-F_interaction}
\end{equation}
where $J=30.4$~krad/s and $r_{jk}$ is the distance between sites $j$ and $k$ (measured in units of lattice constant).
The presence of $^{31}$P nuclear spins-1/2 leads to additional Ising interactions, $H_{\textrm{FP}}=\sum_{j,k}J^{\textrm{FP}}_{jk} S_z^j I_z^k/r^3_{jk}$, where $I_z^k$ is the spin operator of $^{31}$P, $J^{\textrm{FP}}_{jk}$ includes the angular dependence of the dipole-dipole coupling~\cite{SM}.
Crucially, the $^{31}$P nuclear spins are randomly polarized at room temperature and their interaction strength is significantly weaker than both $H_{\textrm{FF}}$ and $H_{\textrm{FP}}$; to this end, $I_z^k$ can be approximated as a scalar random variable, which effectively plays the role of a static, on-site disorder field for the $^{19}$F spins:
\begin{equation}
H_{\textrm{dis}}=\sum_j w_j S_z^j,
\label{eq:random_field}
\end{equation}
where $w_j$ is drawn from a Gaussian distribution with an estimated width of 7 krad/s.

In order to probe the infinite temperature transport of spin and energy in our system, one must measure autocorrelation functions of the form $\sim \mathrm{Tr}[S^j_z(t)S^j_z(0)]$.
To do so, we begin by evolving a weakly polarized thermal state $\rho_0\propto(\mathbb{I}+\epsilon \sum_j S_z^j)$ into a target initial state $\rho\propto\mathbb{I}+\epsilon\mathcal{O}_\rho$.
Next, we evolve this initial state under a desired Hamiltonian $H$ for a time $t$, yielding $\rho(t)=e^{-iHt}\rho e^{iHt}$.
Finally, we measure a tunable observable, $\mathcal{O}_\textrm{m}$; in practice, via RF pulses, this observable is mapped onto the magnetization along the $x$-axis, $\mathcal{M} = \sum_j S_x^j$, which we directly read out via an inductive measurement. 
The resulting signal is equivalent to the infinite temperature correlation function,  $\mathrm{Tr}[\mathcal{O}_\rho(t)\mathcal{O}_\textrm{m}(0)]$. 
Clearly, if $\mathcal{O}_\rho$ and $\mathcal{O}_\textrm{m}$ are  translationally-invariant, the measured signal contains non-local correlations between all pairs of spins, e.g.~$\sum_{jk}\mathrm{Tr}[ S_z^j(t) S_z^k(0)]$. 



To access local correlation functions, such as the spin survival probability~\cite{hunt1956some,zu2021emergent}, we prepare initial states and measure observables such that the spin-polarization at different sites is uncorrelated and  averages to zero.
An exemplary goal is to prepare and measure the random Zeeman state given by $\mathcal{O}_\rho=\sum_j \alpha_j S^j_z(t)$, where $\alpha_j$ are independent and identically distributed random variables with zero average.
This would immediately enable the measurement of single-site autocorrelations since  $\sum_{j,k}\langle \alpha_j \alpha_k\rangle\mathrm{Tr}[S^j_z(t)S^k_z(0)]\propto\sum_{j,k}\delta_{jk}\mathrm{Tr}[S^j_z(t)S^k_z(0)] = \sum_{j} \mathrm{Tr}[S^j_z(t)S^j_z(0)]$. 

Let us now describe our disorder-based experimental protocol for preparing $\mathcal{O}_\rho$  (Fig.~\ref{fig1}C). 
First, we rotate the thermal polarization to the x-axis, initializing a state  
$\propto(\mathbb{I}+\epsilon \sum_j S_x^j)$.
Then, we evolve under $H_{\textrm{dis}}$ for a time $\tau$, such that the excess magnetization of each spin is oriented along a random direction in the xy-plane. 
In order to ensure that the time evolution during $\tau$ is generated only by $H_{\textrm{dis}}$, we utilize concatenated WAHUHA sequences to dynamically decouple $H_\textrm{FF}$~\cite{waugh1968approach}.
%
Next, we employ phase cycling  to project the random polarization of each spin onto the $y$-axis. 
A final RF pulse returns the polarization along $z$, and we obtain $\mathcal{O}_{\rho}=\sum_j \alpha_j S^j_z$, with $\alpha_j=\sin(w_j\tau)$~\cite{SM}. 
A similar strategy can be used to enable a measurement of $\mathcal{O}_{\textrm{m}}=\sum_j \alpha_j S^j_z$.
%
%
In particular, just prior to the final inductive measurement of $\mathcal{M}$, we refocus the random state back to a uniform magnetization by applying the disorder field again.
An analogous approach can be used to detect autocorrelations of two-site observables, such as the local energy density. 
We first use the Jeener-Broekaert pulse pair~\cite{jeener1967nuclear} to create a homogeneous two-body correlated initial state  $\propto\mathbb{I}+ \epsilon \sum_j(S_x^jS_y^{j+1}+S_y^jS_x^{j+1})$
\footnote{Here we assume nearest-neighbor coupling for representation simplicity, but the results also hold with $1/r^3$ long-range coupling \cite{SM}.}.
Evolution under the disordered field and phase cycling yields the random double-quantum (DQ) state with $\mathcal{O}_{\rho}=\sum_j \alpha'_j(S_x^jS_x^{j+1}-S_y^jS_y^{j+1})$ where $\alpha'_j=\sin(w_j\tau+w_{j+1}\tau)$ and $\langle \alpha'_j \alpha'_k\rangle\propto\delta_{jk}$ for large $\tau$. 
An additional $\pi/2$-pulse naturally realizes $\mathcal{O}_{\rho}=\sum_j \alpha'_j(S_y^jS_y^{j+1}-S_z^jS_z^{j+1})$.
We note that linear combinations of these two initial states allow us to reconstruct  all of the subsequent operators we will consider. 

We can carefully characterize the initial state preparation, focusing on two properties: (i) demonstrating that $\mathcal{O}_{\rho}$ has support only on the desired operators
and (ii)  confirming that $\sum \alpha_j=0$ and $\sum \alpha'_j=0$.
%
For the first property, we measure $I(\phi,\theta,\gamma)=\mathrm{Tr}[U_r(\phi,\theta,\gamma) \mathcal{O}_\rho U_r^\dagger(\phi,\theta,\gamma)\mathcal{O}_\rho ]$ for various $\{\phi,\theta,\gamma\}$, where $U_r=\otimes_j e^{-i\gamma S_z^j}e^{-i\theta S_y^j}e^{-i\phi S_z^j}$.
From $I(\phi,\theta,\gamma)$ we can obtain the principal components, $\mathcal{T}_\mu$, of the random observable up to a rotationally-invariant component, $\mathcal{O}_\rho=\sum_\mu d_\mu \mathcal{T}_\mu$, where $d_\mu$ are independent random variables satisfying $\mathbb{E}(d_\mu d_\nu)=\lambda_\mu \delta_{\mu\nu}$, with $\lambda_\mu$ being the eigenvalues of the correlation matrix in descending  order; note that the principal components $\mathcal{T}_\mu$ are orthonormal, $\mathrm{Tr}(\mathcal{T}_\mu\mathcal{T}_\nu^\dagger)=2^L \delta_{\mu\nu}$. The first two principal components  are shown in Fig.~\ref{fig2}A,B, confirming our preparation of the random Zeeman state $\mathcal{O}_{\rho}=\sum_j\alpha_j\sigma_z^j$ and the random DQ state $\mathcal{O}_{\rho}=\sum_j\alpha_j'(\sigma_x^j\sigma_x^{j+1}-\sigma_y^j\sigma_y^{j+1})$ with high fidelity.
As $I(\phi,\theta,\gamma)$ is quadratic in $\mathcal{O}_\rho$, it does not contain information about the sign of the individual random coefficients $\alpha_j$ and $\alpha_j'$.
Therefore, for the second property, we measure the overlap of a random state $\mathcal{O}_\rho$ with its corresponding homogeneous state. As depicted in Fig.~\ref{fig2}C,  the overlap quickly decays to zero as a function of the preparation time, indicating that for sufficient time-evolution under the disordered field, one naturally realizes $\mathbb{E} \alpha_j = \mathbb{E} \alpha'_j=0$.

\begin{figure*}
    \centering
    \includegraphics[width=0.65\textwidth]{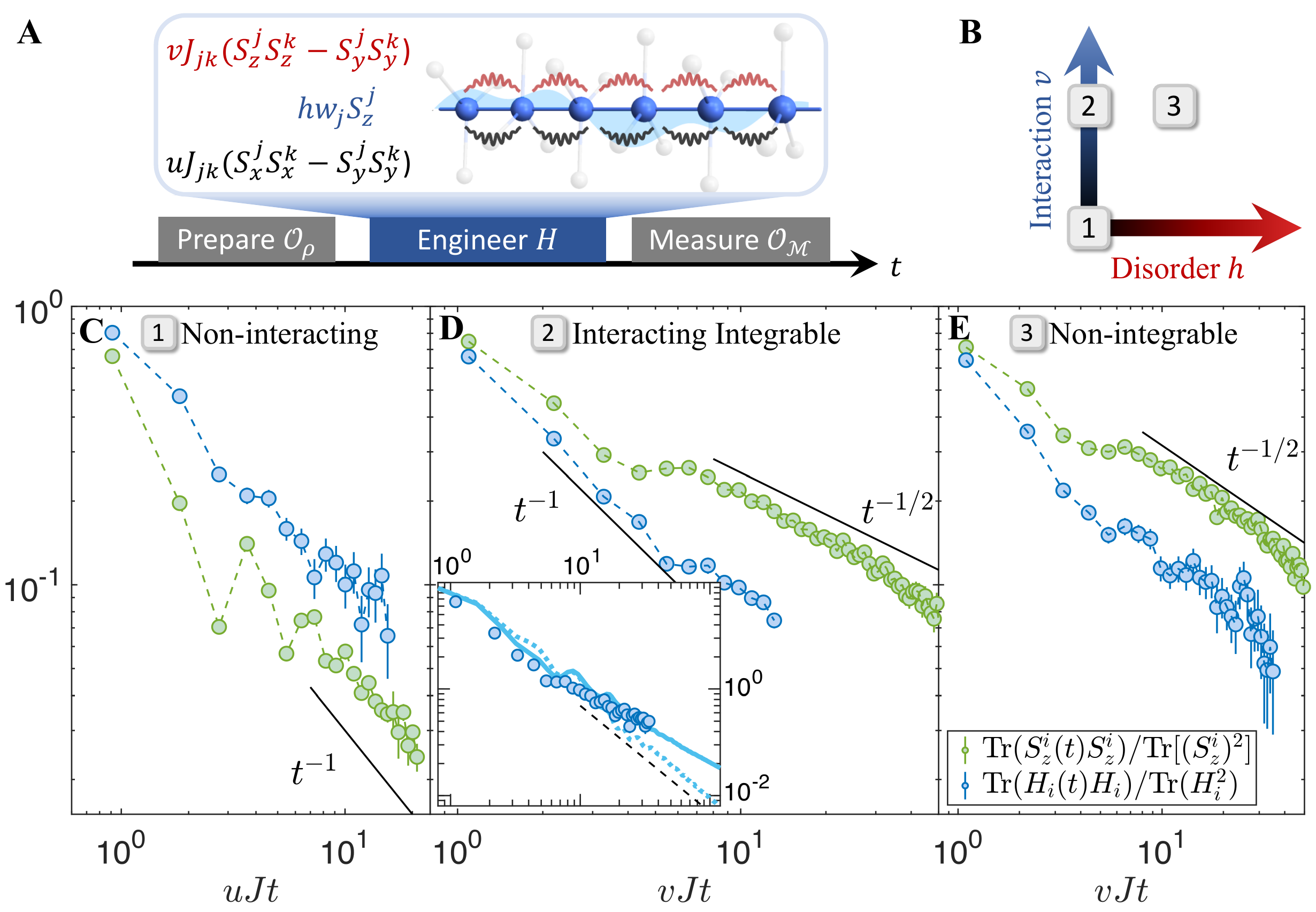}
    \caption{\textbf{Observing different universality classes of hydrodynamics.}
    (\textbf{A}) Utilizing Floquet Hamiltonian engineering techniques, we can independently tune the strengths of two types of interactions (red and black wavy lines) and random on-site field (cyan shaded area). Different combinations of the three terms result in distinct classes of Hamiltonian: 1) Non-interacting, 2) interacting integrable, and 3) non-integrable. 
    (\textbf{B}) Parameter space of the effective Hamiltonian in Eq.~\ref{eq:H} with varying $v,h$ and fixed $u$. 
    (\textbf{C})(\textbf{D})(\textbf{E}) Local autocorrelations of spin and energy in non-interacting, interacting integrable, and non-integrable systems. 
    All these autocorrelations follow power-law decay $t^{-1/z}$, in which the value of the dynamical exponent $z$ distinguishes between different universality classes. 
    Inset of (\textbf{D}): The measured energy autocorrelation (dark blue dots) agrees better with numerical simulation with next-nearest-neighbor coupling (solid curve) than without it (light blue dots), suggesting that the small deviation from ballistic transport at late times is due to the long-range interaction that weakly breaks the integrability of the system.
    Note that we normalize the local autocorrelations by the corresponding global autocorrelations, except the spin autocorrelation in the non-interacting case as we do not have access to the collective conserved quantity $\rho=\sum_j (-1)^j S_z^j$. }
    \label{fig3}
\end{figure*}

\emph{Probing emergent spin and energy hydrodynamics}---
Having verified our initial state preparation, we now turn to exploring the infinite-temperature transport of both spin and energy for three distinct classes of Hamiltonians (Fig.~\ref{fig3}B): (1)  non-interacting integrable, (2)  interacting integrable, and (3) non-integrable. 
Utilizing Floquet engineering, we build each of these Hamiltonians from the native dipolar interaction~\cite{haeberlen1968coherent,peng2021deep}.
In particular, our experiments enable the realization of the following tunable model,
\begin{equation}
\begin{split}
        H=&u\sum_{j<k}\frac{J}{r^3_{jk}}\left(S_x^jS_x^{k}-S_y^jS_y^{k}\right)\\
        +&v\sum_{j<k}\frac{J}{r^3_{jk}}\left(S_z^jS_z^{k}-S_y^jS_y^{k}\right)\\
        +&h\sum_jw_jS_z^j,
\end{split}\label{eq:H}
\end{equation}
where the coefficients $\{u,v,h\}$ can be independently controlled.
For $\{v,h\} = 0$ and restricting to nearest-neighbor couplings (i.e.~truncating the long-range dipolar tail), the resulting XY model is integrable and free (case 1).
Upon adding non-zero $v$, the model remains integrable, but becomes interacting (case 2). 
Finally, the addition of a weak on-site random field, $h$, causes the model to generically become non-integrable (case 3). 
We note that the long-range nature of the dipolar interaction renders $H$ generically \emph{non-integrable} for all of the above cases. However, our hope is that signatures of integrability will be present in the dynamics at short times; as we will see below, this is indeed borne out by the data.

These three different  universality classes can be distinguished by the dynamical exponent, $z$, associated with their spin and energy transport.
Crucially, $z$ can be directly measured via the power-law decay of the autocorrelation function $\sim t^{-1/z}$, with $z=1$ corresponding to ballistic motion, while $z=2$ corresponds to diffusion.

Let us begin with case 1. 
We tune $\{u,v,h\}=\{0.5,0,0\}$ and measure the spin-spin autocorrelation function (Fig.~\ref{fig3}C, green) and the energy autocorrelation function (Fig.~\ref{fig3}C, blue).
Both exhibit late-time power-laws consistent with $z=1$, in agreement with the expectation that quasiparticles propagate  ballistically in a non-interacting, integrable model. 
For case 2, we tune our system to  $\{u,v,h\}=\{-0.15,0.3,0\}$.
Intriguingly, we find that spin transports diffusively while  energy transports ballistically (Fig.~\ref{fig3}D).
This phenomenon owes to the existence of stable spinless quasiparticles and is a central feature of infinite temperature transport in the so-called XXZ model~\cite{grabowski1995structure,zotos1997transport,klumper2000thermodynamics,sakai2003non,prosen2009matrix,steinigeweg2009density,vznidarivc2011spin,karrasch2014real,de2019diffusion,gopalakrishnan2019kinetic}.
Finally, for case 3, we set $\{u,v,h\}=\{-0.15,0.3,0.23\}$ and observe that  both spin and energy transport diffusively (Fig.~\ref{fig3}E), consistent with a generic non-integrable model~\cite{spohn2012large,friedman2020diffusive}. 

\begin{figure*}
    \centering
    \includegraphics[width=0.65\textwidth]{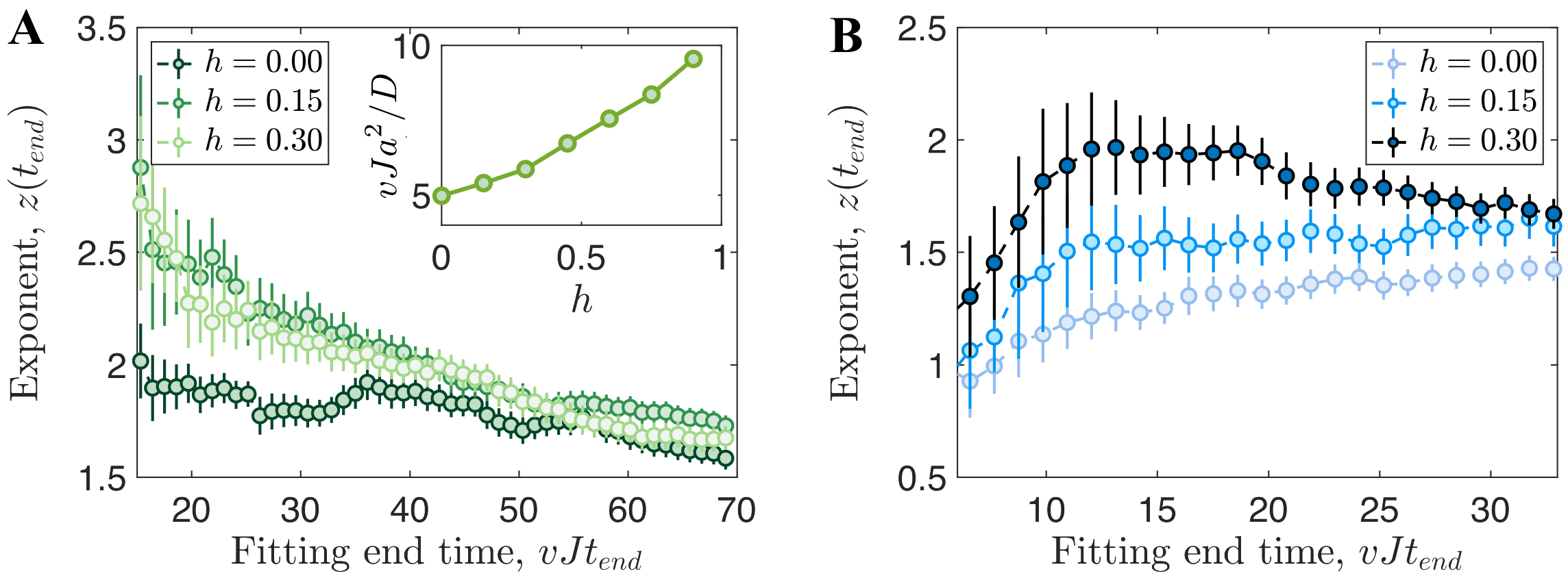}
    \caption{\textbf{Finite-time effect of transport in the presence of on-site random field.}
    Extracted dynamical exponent of spin (\textbf{A}) and energy (\textbf{B}) transport in different fitting time windows.  
    The fitting window starts after the transient dynamics [$Jt_{start}=7.7$ for (\textbf{A}) and $Jt_{start}=2.2$ for (\textbf{B})]; note that the qualitative features we observed do not depend on this specific choice. 
    Inset of (\textbf{A}): Inverse of the diffusion constant $D$ extracted by fitting the data from $Jt_{start}=7.7$ to $Jt_{end}=60.0$, with  $v=0.3$ and $a=3.442\AA$ the FAp lattice constant.}
    \label{fig4}
\end{figure*}

Two remarks are in order. First, the energy transport data in case 2 exhibit a weak deviation from ballistic transport at the longest times explored in the experiment (inset, Fig.~\ref{fig3}D). In order to understand the origin of this deviation,  we numerically compute the energy autocorrelation function using density matrix truncation, with and without long-range couplings~\cite{ye2020emergent}. The agreement between our experiment and numerics in the former case suggests that the observed deviation results from the weak breaking of integrability associated with the long-range couplings. 
Second, by tuning the disorder strength during the evolution, we can controllably break integrability and access the nonintegrable regime on the experimental timescale.
In Fig.~\ref{fig4}, we measure the energy and spin transport as we tune $h$ from 0 to 0.3.
We extract $z$ using different time windows of the   autocorrelation function, starting at  $t_{start}=7.7/J$ and ending at a variable $t_{end}$. 
For the spin transport (Fig.~\ref{fig4}A), after an initial transient, all of the models exhibit a $z=2$ at intermediate times. At the latest times, the inter-chain couplings begin to play a role, causing a decrease in $z$. 
Meanwhile, for the energy transport at $h=0$, $z$ remains close to its initial ballistic value for all times.
However, for $h=0.3$, the system reaches a diffusive exponent ($z=2$) at intermediate times before exhibiting a weak decrease (possibly owing to interchain couplings). 

In summary, our results introduce a novel method to probe local spin and energy transport in solid-state spin ensembles. 
Our technique leverages the intrinsic disorder in such systems and requires only collective control. 
Using our method, we demonstrate the observation of ballistic and diffusive hydrodynamics in a variety of one-dimensional spin chains.
Looking forward, our protocol could be used to explore the existence of subdiffusion near the many-body localization transition or the emergence of superdiffusion with long-range interactions \cite{lucioni2011observation,vosk2015theory,potter2015universal,sahay2021emergent,agarwal2015anomalous,zaburdaev2015levy,joshi2022observing}.
Finally, in addition to two-point correlation functions, which were the focus of the present work, our protocols can naturally be generalized to four-point, out-of-time-ordered correlations, and thus used to probe many-body quantum information scrambling \cite{nahum2018operator,von2018operator,rakovszky2018diffusive,khemani2018operator,xu2020accessing,xu2019locality,sahu2019scrambling,schuster2021many,li2017measuring,landsman2019verified,blok2021quantum,wei2019emergent}.

\textit{Note added}: after the completion of this project, we became aware of related work~\cite{Martin22}, which develops similar techniques and applies them to study local thermalization.

\textbf{Acknowledgments}\\
We thank Chandrasekhar Ramanathan, Hengyun Zhou, Martin Leigh, Nathaniel Leitao, Francisco Machado, Jack Kemp, Joel Moore and Mikhail Lukin for helpful conversations.
This work was supported in part by the National Science Foundation under Grants No. PHY1915218.
P.P. thanks MathWorks for their support in the form of a Graduate Student Fellowship. The opinions and views expressed in this publication are from the authors and not necessarily from MathWorks.\\
B.Y. acknowledges support from the Army Research Office through the MURI program (W911NF-20-1-0136).
N.Y.Y. acknowledges support from the U.S. Department of Energy, Office of Science, through the Quantum
Systems Accelerator (QSA), a National Quantum Information Science Research Center and the David and Lucile Packard foundation.

\textbf{Author Contributions}\\
P.P. designed and performed the experiment with assistance from P.C.. B.Y. and N.Y.Y. performed the numerical and analytical calculations. P.C. supervised the project. All authors worked on the interpretation of the data and contributed to writing the manuscript. \\

\textbf{Competing interests}\\
The authors declare no conflict of interest.
\clearpage

\section{Notations}
For easier representation, we introduce the notations for operators on spin-1/2 chains in Table.~\ref{tab:notations}. 
\begin{table}[h]
\centering
    \begin{tabular}{|c|c|}
        \hline
        Notations & Operators \\
        \hline
        $X$  & $\sum_j S_x^j$\\
        \hline
        $Y$  & $\sum_j S_y^j$\\
        \hline
        $Z$  & $\sum_j S_z^j$\\
        \hline
        $D_x$  & $\frac{1}{2}\sum_{j<k}(2 S_x^j S_x^{k}-S_y^j S_y^{k}-S_z^j S_z^{k})/|k-j|^{3}$\\
        \hline
        $D_y$  & $\frac{1}{2}\sum_{j<k}(2 S_y^j S_y^{k}-S_x^j S_x^{k}-S_z^j S_z^{k})/|k-j|^{3}$\\
        \hline
        $D_z$  & $\frac{1}{2}\sum_{j<k}(2 S_z^j S_z^{k}-S_x^j S_x^{k}-S_y^j S_y^{k})/|k-j|^{3}$\\
        \hline
        $DQ_x$  & $\sum_{j<k}(S_y^j S_y^{k}-S_z^j S_z^{k})/|k-j|^{3}$\\
        \hline
        $DQ_y$  & $\sum_{j<k}(S_z^j S_z^{k}-S_x^j S_x^{k})/|k-j|^{3}$\\
        \hline
        $DQ_z$  & $\sum_{j<k}(S_x^j S_x^{k}-S_y^j S_y^{k})/|k-j|^{3}$\\
        \hline
    \end{tabular}
    \caption{Notations for operators used in this paper.}
\label{tab:notations}
\end{table}

We use prefix $r$ to denote random operators. For example, $rZ=\sum_j \alpha_j S_z^j$, $rDQ_Z=\sum_{j}\alpha_{j}'(S_x^j S_x^{j+1}-S_y^j S_y^{j+1})$, with $\alpha_j, 
\alpha_{j}'$ independent and identically distributed random variables with zero mean, $\mathbb{E}(\alpha_j)=0,\mathbb{E}(\alpha_{j}')=0,\mathbb{E}(\alpha_j\alpha_k)\propto \delta_{jk}, \mathbb{E}(\alpha_{j}'\alpha_{k}')\propto \delta_{jk}$.
\section{Experimental system}
The sample in the experiment is a single crystal of fluorapatite (FAp) with formula Ca$_5$(PO$_4$)$_3$F. The most abundant isotopes of F and P have 1/2 nuclear spin, while the most abundant isotopes of Ca and O have zero nuclear spin. Fluorapatite is a hexagonal mineral with space group P\(6_3/m\), where the \(^{19}\)F spin-1/2 nuclei form linear chains along the \(c\)-axis. Each fluorine spin in the chain is surrounded by three equidistant \(^{31}\)P spin-1/2 nuclei. 
The sample we used is a cut from a natural crystal of approximate dimensions 3 mm$\times$3 mm$\times$2 mm.
The sample is placed at room temperature inside a superconducting magnet producing a uniform $B=7$~T field. The total Hamiltonian of the system is given by
\begin{equation}
H_\mathrm{tot}=\omega_F \sum_k S_z^k+\omega_P \sum_\kappa s_z^\kappa+H_\mathrm{dip}
\label{eq:Hamtot}	
\end{equation}
The first two terms represent the Zeeman interactions of the F($S$) and P($I$) spins, respectively, with frequencies $\omega_F=\gamma_FB\approx (2\pi)282.37$ MHz and $\omega_P=\gamma_PB=(2\pi)121.51$ MHz, where $\gamma_{F/P}$ are the gyromagnetic ratios. The last term represent the natural magnetic dipole-dipole interaction among the spins, given by
\begin{equation}
\begin{aligned}
 H_\mathrm{dip}&=H_\mathrm{FF}+H_\mathrm{FP}+H_\mathrm{PP}\\
 &=\sum_{j<k}\frac{\hbar\gamma_j\gamma_k}{|\vec r_{jk}|^3}\left[\vec S_j\cdot\vec S_k-\frac{3(\vec S_j\cdot\vec r_{jk})\,(\vec S_k\cdot\vec r_{jk})}{|\vec r_{jk}|^2}\right],
 \end{aligned}
\end{equation}
where $\vec r_{jk}$ is the vector between the $jk$ spin pair. Because the Zeeman interaction is much stronger than dipole-dipole interaction, we can truncate the dipolar Hamiltonian to its energy-conserving part (secular Hamiltonian). We then obtain the homonuclear Hamiltonians
\begin{equation}
 \begin{aligned}
 H_\mathrm{FF}&=\frac{1}{2}\sum_{j<k}J^F_{jk}(2 S_z^j S_z^{k}- S_x^j S_x^{k}- S_y^j S_y^{k}) \\ H_\mathrm{PP}&=\frac{1}{2}\sum_{\lambda<\kappa}J^P_{\kappa\lambda}(2I_z^\lambda I_z^{\kappa}-I_x^\lambda I_x^{\kappa}-I_y^\lambda I_y^{\kappa})
 \end{aligned}
\end{equation}
and the heteronuclear interaction between the $F$ and $P$ spins,
\begin{equation}
 H_\mathrm{FP}=\sum_{k,\kappa} J^\mathrm{FP}_{k,\kappa}S_z^kI_z^\kappa,
\end{equation}
with $J_{jk}=\hbar\gamma_j\gamma_k\frac{1-3\cos(\theta_{jk})^2}{|\vec r_{jk}|^3}$, where $\theta_{jk}$ is the angle between the vector $\vec r_{jk}$ and the magnetic field $z$-axis. We align the $c$-axis to the 7~T magnetic field. The maximum values of the couplings (for the closest spins) are given respectively by $J=J^\mathrm{FF}=-32.76$ krad s$^{-1}$, $J^\mathrm{PP}=1.20$ krad s$^{-1}$ and $J^\mathrm{FP}=6.12$ krad s$^{-1}$. Since the coupling between $^{31}$P spins is much weaker than the others, dynamics of $^{31}$P spins can be ignored for short time and $I$ only provides a static random field. In addition, as the temperature is much higher than the Zeeman energy, each $^{31}$P spin is randomly polarized with negligible correlation between different $^{31}$P spins. As a result, $H_\mathrm{FP}$ can be viewed as an on-site disordered field for $^{19}$F spins.
\begin{equation}
 H_\mathrm{dip}=\frac{1}{2}\sum_{j<k}J_{jk}(2 S_z^j S_z^{k}-S_x^j S_x^{k}-S_y^j S_y^{k})+\sum_j w_j S_z^j,
\end{equation}
where $w_j=\sum_{\kappa}J_{j,\kappa}^{FP} I_z^\kappa$ is a random number.

The dynamics of this complex 3D many-body system can be mapped to a much simpler, quasi-1D system. First, when the crystal is oriented with its $c$-axis parallel to the external magnetic field the coupling of fluorine spins to the closest off-chain fluorine spin is $\approx40$ times weaker, while in-chain, next-nearest neighbor couplings are $8$ times weaker. 
Previous studies on these crystals have indeed observed dynamics consistent with spin chain models, and the system has been proposed as solid-state realizations of quantum wires~\cite{cappellaro2007simulations,cappellaro2011coherent,ramanathan2011experimental}. This approximation of the experimental system to a 1D, short-range system, although not perfect, has been shown to reliably describe experiments for relevant time-scales~\cite{rufeil2009effective,zhang2009nmr}. The approximation breaks down at longer times, with a convergence of various effects: long-range in-chain and cross-chain couplings, as well as pulse errors in the sequences used for Hamiltonian engineering. In addition, the system also undergoes spin relaxation, although on a much longer time-scale ($T_1=0.8~$s for our sample).

\subsection{Hamiltonian Engineering}\label{app:Ham}
We engineer a large variety of Hamiltonians by periodically applying rf pulse trains to the natural dipolar Hamiltonian that describes the system. The periodically driven (Floquet) system is described by Floquet Hamiltonian that can be different from the natural Hamiltonian. 
Floquet Hamiltonian can be perturbatively calculated using the Average Hamiltonian Theory (AHT~\cite{haeberlen1968coherent}). 
The dynamics is induced by the total Hamiltonian \(H=H_\text{dip}+H_\text{rf}\), 
where \(H_\text{dip}=\frac{1}{2}\sum_{j<k}J_{jk}(2 S_z^j S_z^{k}-S_x^j S_x^{k}-S_y^j S_y^{k})+\sum_j w_j S_z^j\) is the system Hamiltonian, 
and \(H_\text{rf}(t)\) is the external Hamiltonian due to the rf-pulses. 
The density matrix \(\rho\) evolves under the total Hamiltonian according to \(\dot\rho=-i[H,\rho]\). 
We study the dynamics into a convenient interaction frame, defined by \(\rho'={U_\text{rf}}^{\dagger}\rho U_\text{rf}\), where \(U_\text{rf}(t)=\mathcal{T}\exp[-i\int_0^t H_\text{rf}(t') dt']\) and \(\mathcal{T}\) is the time ordering operator. 
In this \textit{toggling} frame, \(\rho'\) evolves according to \(\dot{\rho}'=-i[H(t),\rho']\), where \(H(t)={U_\text{rf}}^{\dagger}H_\text{dip} U_\text{rf}\). 
Since \(U_\text{rf}\) is periodic, \(H(t)\) is also periodic with the same period $\tau$, and gives rise to the Floquet Hamiltonian, $H_F$, as \(U(\tau)=\exp[-i H_F \tau]\). 
Note that if the pulse sequence satisfies the condition \(U_\text{rf}(\tau)=1\), the dynamics of \(\rho\) and \(\rho'\) are identical when the system is viewed stroboscopically, i.e., at integer multiples of \(\tau\), where the toggling frame coincides with the (rotating) lab frame. 

In this work, we engineer the nearly integrable double quantum Hamiltonian using the 8-pulse sequence introduced in~\cite{Yen83}. For the interacting models, we use a 16-pulse sequence. The basic building block is given by a 4-pulse sequence originally developed to study multiple quantum coherence~\cite{kaur2012initialization,yen1983multiple}.
We denote a generic 4-pulse sequence as \(P(\tau_1,{\bf n}_1,\tau_2,{\bf n}_2,\tau_3,{\bf n}_3,\tau_4,{\bf n}_4,\tau_5)\), where \({\bf n}_j\) represents the direction of the \(j\)-th \(\pi/2\) pulse, and \(\tau_j\)'s the delays interleaving the pulses. In our experiments, the \(\pi/2\) pulses have a width \(t_w\) of typically 1.02 \(\mu\)s. \(\tau_j\) starts and/or ends at the midpoints of the pulses. In this notation, the 16-pulse sequence can be expressed as
\begin{widetext}
\begin{gather*}
P(\tau_1,{\bf x},\tau_2,{\bf y},2\tau_3,{\bf y},\tau_2',{\bf x},\tau_1')P(\tau_1',{\bf x},\tau_2,{\bf y},2\tau_3',{\bf y},\tau_2',{\bf x},\tau_1)P(\tau_1,{\bf \overline{x}},\tau_2',{\bf \overline{y}},2\tau_3',{\bf \overline{y}},\tau_2,{\bf \overline{x}},\tau_1')P(\tau_1',{\bf \overline{x}},\tau_2',{\bf \overline{y}},2\tau_1,{\bf \overline{y}},\tau_2,{\bf \overline{x}},\tau_1)
\end{gather*}
where \(\{{\bf \overline{x}},{\bf \overline{y}}\}\equiv \{{\bf -x},{\bf -y}\}\). The delays are given by
\begin{gather*}
\begin{aligned}
\tau_1&=\tau_0(1+c-v+w), \quad
\tau_2=\tau_0(1+b-u+v), \quad
\tau_3=\tau_0(1-a+u-w),\\
\tau_1'&=\tau_0(1-c-v+w), \quad
\tau_2'=\tau_0(1-b-u+v), \quad
\tau_3'=\tau_0(1+a+u-w),\\
\end{aligned}
\end{gather*}
where \(\tau_0\) is 5 \(\mu\)s in this paper. To the second order Magnus expansion, the above sequence realizes the Floquet Hamiltonian 
\begin{equation*}
    H_{F}=\frac{1}{2}\sum_{j<k}J_{jk}\left[(u-w)S_x^j S_x^{k}+(v-u)S_y^j S_y^{k}+(w-v)S_z^j S_z^{k}\right]+\frac{1}{3}\sum_j w_j (aS_x^j+bS_x^j+cS_x^j).
\end{equation*}
\end{widetext}
The cycle time \(t_c\), defined as the total time of the sequence, is given by \(\tau=24\tau_0\). \(u,v,w,a,b,c\) are dimensionless adjustable parameter, and is restricted such that none of the inter-pulse spacings becomes negative.

The method above can be applied more broadly to engineer desired Hamiltonians $ H_{des}$ using only collective rotations of the spins applied to the naturally occurring Hamiltonian, $H_{nat}$. The engineered Hamiltonian is obtained by piece-wise constant evolution under-rotated versions of the natural Hamiltonian under the condition
\(
\sum_k R_k H_{nat} R_k^\dag = H_{des},
\)
where $R_k$ are collective rotations of all the spins, which achieves the desired operator to first order in a Magnus expansion. Symmetrization of the sequence can further cancel out the lowest order correction. 
Using only collective pulses limits which Hamiltonians can be engineered, due to symmetries of the natural Hamiltonian and the action of collective operators. For typical two-body interactions of spin-1/2, an efficient tool to predict which Hamiltonians are accessible is to use spherical tensors~\cite{ajoy2013quantum}.

\subsection{\textit{Ab initio} calculation of disordered field}
\begin{figure*}[!htp]
    \centering
    \includegraphics[width=0.9\textwidth]{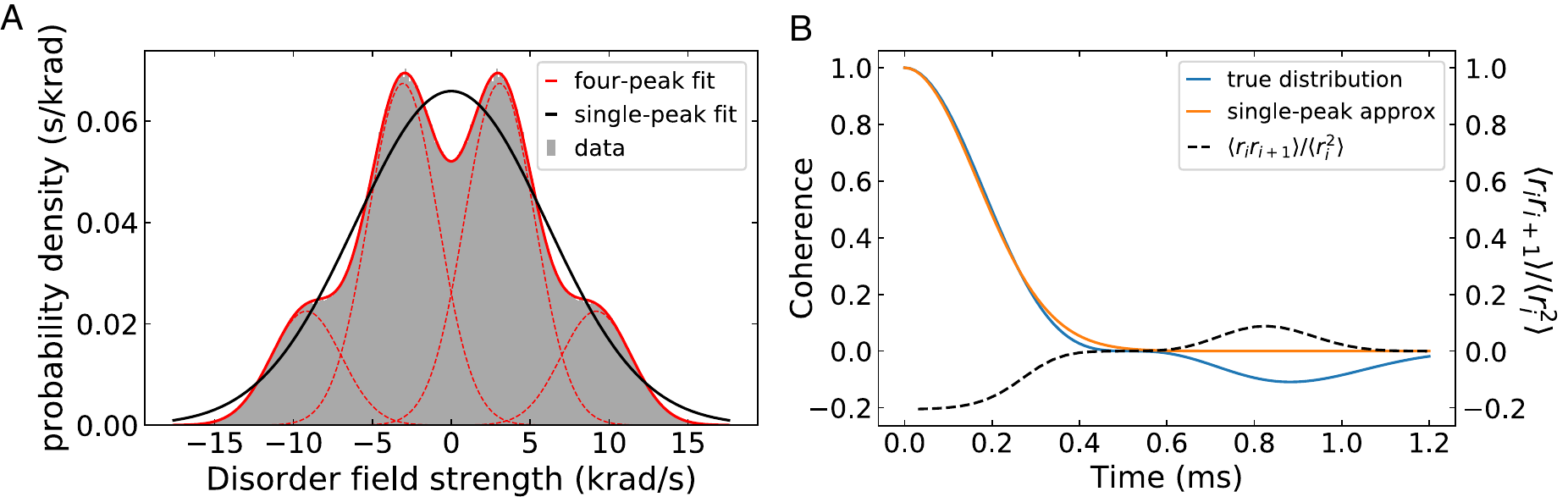}
    \caption{Disordered on-site field generated by $^{31}$P. (A) Numerical calculation of distribution of the on-site field strength. (B) Left axis: Decoherence profile generated by the calculated distribution of on-site field and the single-peak Gaussian approximation. Right axis: Statistical correlation between the random amplitudes of local observables on two closest $^{19}$F. As the coherence approaches zero, the statistical correlation also vanishes. }
    \label{fig:disorderfield}
\end{figure*}

The disordered magnetic fields on the $^{19}$F originates from the $^{19}$F-$^{31}$P interaction and the random orientation of $^{31}$P. 
This picture allows us to directly calculate the distribution of the disordered field strength. 
In particular, we compute the interaction strength between $^{19}$F and its several closest neighboring $^{31}$P based on their relative position~\cite{Comodi01} and gyromagnetic ratio. 
Then we assume each $^{31}$P points along $+z$ or $-z$ direction with the same probability, which effectively applies a magnetic field on $^{19}$F along the corresponding direction and with the strength given by $^{19}$F-$^{31}$P interaction. 
Summing up the contribution from all $^{31}$P gives the total strength of the disorder field. 
In Fig.~\ref{fig:disorderfield}A, we include the 45 closest $^{31}$P and observe a smooth enough distribution of the field strength. 
The distribution is perfectly fitted by a sum of four Gaussian distributions with the centers at $\pm \frac{1}{2}J^{FP}$ and $\pm \frac{3}{2}J^{FP}$ and the height ratio of $3:3:1:1$. 
Interestingly, even a single Gaussian function can still capture the distribution reasonably well, so we simply use a Gaussian distribution for simplicity in our numerical simulation. 
This can be qualitatively justified by comparing the computed $T_2$ decay profiles of the Gaussian distribution and the true distribution (Fig.~\ref{fig:disorderfield}B). 

Since the disorder fields on different $^{19}$F originates from the same $^{31}$P bath, they inevitably have some statistical correlation. 
Here, we evaluate the correlationof the disordered fields on two neighboring $^{19}$F, $\frac{\langle w_j w_{j+1}\rangle}{\langle w_j^2\rangle}\approx -0.2$. 
Nevertheless, $\langle \alpha_j \alpha_{j+1}\rangle = \langle \mathrm{sin}(w_j\tau)\; \mathrm{sin}(w_{j+1}\tau)\rangle\approx 0$ for $\tau > T_2$ (Fig.~\ref{fig:disorderfield}), satisfying the condition required for our protocol to measure local autocorrelation (see Sec.~\ref{sec:spatialCorr} for more discussion). 

\section{Details about random state creation and detection}
\label{sec:DetailSeq}
\begin{figure*}[!htp]
    \centering
    \includegraphics[width=0.97\textwidth]{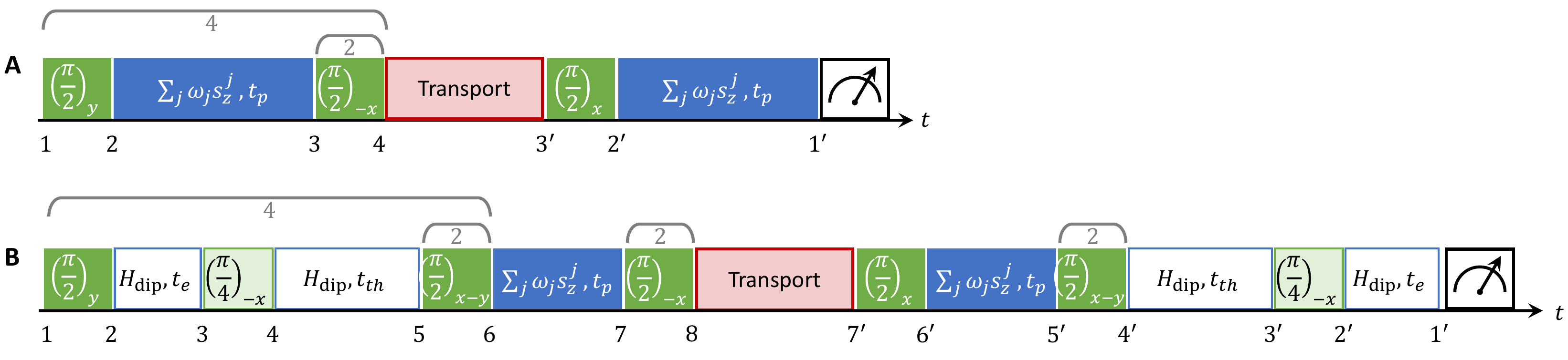}
    \caption{Detailed sequence of the creation and detection of random Zeeman state (a) and double quantum state (b). Phase cycling is mark by brackets above the sequence, with the number under the bracket labels number of experiments with that phase cycling. For easier reference, we label different part of the sequence by numbers listed in the bottom.}
    \label{fig:DetailSeq}
\end{figure*}
The sequences we use to study transport contains 3 parts -- state engineering, transport Hamiltonian engineering and observable engineering. The total unitary propagator $U_{tot}$ is thus a product of 3 unitaries, $U_{tot}=U_{\rho}U_H U_\mathcal{O}$ and the NMR signal is then $\mathrm{Tr}(U_{tot}\rho_0U_{tot}^\dagger\mathcal{O}_0)$.
Figure~\ref{fig:DetailSeq} shows the sequences we use to create, evolve, and detect random states. In the following we explain the sequence step by step.

\paragraph*{Random Zeeman state}
\begin{itemize}
\item[1:] The state is initially the high-temperature equilibrium state, $\rho\propto \mathbb{I}+\epsilon \rho_1$, with -$\rho_1=Z=\sum_jS_z^j$. Since the identity does not evolve nor give rise to signal, in the following we only report the dynamics of the deviation $\rho_1$.
\item[2:] $\rho_2=X$ thanks to a $\pi/2$ pulse along the y axis.
\item[3:] The disorder Hamiltonian is engineered with two concatenated WAHUHA sequences (WAHUHA8, eight $\pi$/2 pulses along x, y, -y, -x, -x, -y, y, x)~\cite{waugh1968approach}. The WAHUHA8 sequence cancels the $^{19}$F couplings, yielding the average Hamiltonian  $\sum_j \omega_j S_z^j$, with $\omega_j=(1/3)w_j$. The total sequence length is 60$\mu$s. All the disorder Hamiltonians in Fig.~\ref{fig:DetailSeq} are engineered by this sequence. The state becomes $\rho_3=\sum_j \cos(\omega_j \tau)S_x^j+\sin(\omega_j \tau)S_y^j$.
\item[4:] To cancel the $S_x$ component, we apply a 2-fold phase cycle over the $\pi/2$ pulse between 3 and 4. That is, we implement the sequence shown in Fig.~\ref{fig:DetailSeq}A twice, one with $(\pi/2)_x$ yielding $\rho_4^{(1)}=\sum_j [\cos(\omega_j \tau)S_x^j+\sin(\omega_j \tau)S_z^j]$, the other with $(\pi/2)_{-x}$ yielding $\rho_4^{(2)}=\sum_j [\cos(\omega_j \tau)S_x^j-\sin(\omega_j \tau)S_z^j]$. Then we take the difference of the signals from the two experiments as the final signal $S=\frac{1}{2}[\mathrm{Tr}(\rho_4^{(1)}(t)\mathcal{O})-\mathrm{Tr}(\rho_4^{(2)}(t)\mathcal{O})]=\mathrm{Tr}(\rho_4(t)\mathcal{O})$, with $\rho_4=\frac{1}{2}(\rho_4^{(1)}-\rho_4^{(2)})=\sum_j \sin(\omega_j \tau)S_z^j$. We further implement a 4-fold phase cycle over all pulses before step 4, by a 4-fold rotation along z. This amounts to implementing  each of the two repetitions discussed above  4 times, with 90$^\circ$ phase increments of all pulses before step 4 each time, and average over the results. Ideally the rotation along z axis should not affect $\rho_4$, but due to experimental imperfections $\rho_4$ also contains some unwanted terms that can be averaged out under this 4-fold phase cycling. There are 8 repetitions in total.
\item[$1'$:] Creating the random observable can be best analyzed by describing the steps from later to earlier times. The final observable is $\mathcal{O}_{1'}=X$.
\item[$2'$:] Evolution under the disorder Hamiltonian (obtained with a WAHUHA8 sequence) yields  $\mathcal{O}_{2'}=\sum_j[\cos(\omega_j \tau)S_x^j-\sin(\omega_j \tau)S_y^j]$.
\item[$3'$:] $\mathcal{O}_{3'}=\sum_j [\cos(\omega_j \tau)S_x^j+\sin(\omega_j \tau)S_z^j]$. If the transport Hamiltonian does not contain a disordered field, then the first term has zero overlap with the density matrix $\rho_4$ and can be neglected, leading to $\mathcal{O}=\sum_j \sin(\omega_j \tau)S_z^j$. If the transport Hamiltonian contains the disordered field, then another 2-fold phase cycling over the $\pi/2$ pulse between step $3'$ and $2'$ is needed to engineer the same observable, similar to step 4 above.
\item[$\bullet$] Signal: The final signal after all phase cycling is $S=\mathrm{Tr}(\rho(t)\mathcal{O})$, with $\rho(0)=\mathcal{O}=\sum_j \sin(\omega_j \tau)S_z^j$. This already looks like the autocorrelation of random Zeeman state with random coefficient $\alpha_j=\sin(\omega_j \tau)$. It is easy to show $\sin(\omega_j \tau)$ has zero mean. We will show $\mathbb{E}[\sin(\omega_j \tau)\sin(\omega_k \tau)]\propto \delta_{jk}$ in Sec.~\ref{sec:properties}.
\item Variations: If we want to measure autocorrelation of random Zeeman state along y axis, we can simply move the transport step after the $\pi/2$ pulse between $3'$ and $2'$. That $\pi/2$ pulse is then considered part of initial state engineering, and the final effective initial state density matrix is $\rho=\sum_j \sin(\omega_j \tau)S_y^j$ and observable is $\mathcal{O}=\sum_j \sin(\omega_j \tau)S_y^j$. Random Zeeman state along x axis can be engineered with 90$^\circ$ rotation of all pulses (except the transport Hamiltonian engineering pulses) along z axis. Random Zeeman state along any other axis can be decomposed in to correlations of random Zeeman states along x, y, z axes.
\end{itemize}

\textit{Random double quantum state}
\begin{enumerate}[label=\arabic*:]
\item  $\rho_1=Z$.
\item  $\rho_2=X$.
\item  $\rho_3=X-i[H_\mathrm{dip},X]t_e+O(t_e^2)=X-\frac{3}{4}\sum_{j<k}J_{jk}(S_z^jS_y^{k}+S_y^jS_z^{k})t_e+\sum_jw_jt_eS_y^j+O(t_e^2)$.
\item $\rho_4=\sum_j S_x^j+\frac{3}{4}\sum_{j<k}J_{jk}( S_z^jS_z^{k}-S_y^jS_y^{k})t_e+\sum_jw_jt_e(S_y^j-S_z^j)/\sqrt{2}+O(t_e^2)$.
\item  The evolution time between step 4 and 5 is very long $|H_\mathrm{dip}t_{th}|\gg 1$ such that state $\rho_4$ thermalizes to $e^{-\beta H_\mathrm{dip}}/\mathrm{Tr}(e^{-\beta H_\mathrm{dip}})$~\cite{deutsch1991quantum,rigol2008thermalization}. The inverse temperature $\beta$ is determined by the energy conservation $\mathrm{Tr}[H_\mathrm{dip}(1-\epsilon \rho_4)]=\mathrm{Tr}(H_\mathrm{dip}e^{-\beta H_\mathrm{dip}})/\mathrm{Tr}(e^{-\beta H_\mathrm{dip}})$. The state after thermalization is still a high temperature state with $\beta=O(\epsilon)$, therefore we have $e^{-\beta H_\mathrm{dip}}/\mathrm{Tr}(e^{-\beta H_\mathrm{dip}})\approx(\mathbb{
I}-\beta H_\mathrm{dip})/\mathrm{Tr}(\mathbb{I})$ and the non-identity part is $\rho_5\propto H_\mathrm{dip}$. The sequence to create dipolar state was first demonstrated in Ref.~\cite{jeener1967nuclear}.
\item  $\rho_6^{(0)}=-\frac{1}{2} D_z+\frac{3}{4}\sum_{j<k}(S_x^j S_y^k+S_y^j S_x^k)/|k-j|^3 +\sum \frac{w_j}{\sqrt{2}} (S_x^j +S_y^j)$, the superscript ${(0)}$ indicates this is the density matrix before phase cycling. The last term can be cancelled by 2-fold phase cycling of the $\pi/2$ pulse between step 5 and 6, i.e. averaging two experiments one with $(\pi/2)_{x-y}$ the other with $(\pi/2)_{y-x}$. The first term can be cancelled by 4-fold phase cycling of all previous pulses, i.e. 
$$
\begin{aligned}\rho_{6}&=\frac{1}{4}[\rho_6^{(0)}-e^{-i(\pi/2) Z} \rho_6^{(0)} e^{i(\pi/2) Z}\\
&\quad+e^{-i\pi Z} \rho_6^{(0)} e^{i\pi Z}\\
&\quad-e^{-i(3\pi/2) Z} \rho_6^{(0)} e^{i(3\pi/2) Z}]\\
&=\frac{3}{4}\sum_{j<k}(S_x^j S_y^k+S_y^j S_x^k)/|k-j|^3,
\end{aligned}
$$
notice the minus sign in first and third line.
\item  $\rho_7=\frac{3}{4}\sum_{j<k}[(S_x^j S_y^k+S_y^j S_x^k)\cos(\omega_j\tau+\omega_k\tau)+(S_y^j S_y^k-S_x^j S_x^k)\sin(\omega_j\tau+\omega_k\tau)]/|k-j|^3$.
\item  A 2-fold phase cycling over $\pi/2$ pulse between step 7 and 8 cancels the first term in $\rho_7$. $\rho_8=\frac{3}{4}\sum_{j<k}(S_z^j S_z^k-S_x^j S_x^k)\sin(\omega_j\tau+\omega_k\tau)/|k-j|^3$.
\end{enumerate}
\begin{enumerate}[label=\arabic*$'$:]
    \item $\mathcal{O}_{1'}=-X$.
\item  $\mathcal{O}_{2'}=-X+i[H_\mathrm{dip},-X]t_e+O(t_e^2)=-X-\frac{3}{4}\sum_{j<k}J_{jk}(S_z^jS_y^{k}+S_y^jS_z^{k})t_e+\sum_jw_jt_eS_y^j+O(t_e^2)$. Notice for observable engineering we consider backward evolution so that the effective Hamiltonian is $-H_{dip}$.
\item  $\mathcal{O}_{3'}=-X-\frac{3}{4}\sum_{j<k}J_{jk}(S_z^jS_z^{k}-S_y^jS_y^{k})t_e+\sum_jw_jt_e(S_y^j+S_z^j)/\sqrt{2}+O(t_e^2)$.
\item  Thermalization, $\mathcal{O}_{4'}\propto -H_\mathrm{dip}$
\item  $\mathcal{O}_{5'}^{(0)}=\frac{1}{2} D_z-\frac{3}{4}\sum_{j<k}(S_x^j S_y^k+S_y^j S_x^k)/|k-j|^3 -\sum \frac{w_j}{\sqrt{2}} (S_x^j +S_y^j)$. The last term can be cancelled by 2-fold phase cycling of the $\pi/2$ pulse between step $5'$ and $4'$. $\mathcal{O}_{5'}=\frac{1}{2} D_z-\frac{3}{4}\sum_{j<k}(S_x^j S_y^k+S_y^j S_x^k)/|k-j|^3$.
\item   $\mathcal{O}_{6'}=\frac{1}{2}\sum_{jk}[S_z^jS_z^k-\frac{1}{2}(S_x^jS_x^k+S_y^jS_y^k)\cos(\omega_j\tau-\omega_k\tau)-\frac{1}{2}(S_x^j S_y^k-S_y^j S_x^k)\sin(\omega_j\tau-\omega_k\tau)]/|k-j|^3+\frac{3}{4}\sum_{j<k}[-(S_x^j S_y^k+S_y^j S_x^k)\cos(\omega_j\tau+\omega_k\tau)+(S_y^j S_y^k-S_x^j S_x^k)\sin(\omega_j\tau+\omega_k\tau)]/|k-j|^3$.
\item   $\mathcal{O}_{7'}=\frac{1}{2}\sum_{jk}[S_y^jS_y^k-\frac{1}{2}(S_x^jS_x^k+S_z^jS_z^k)\cos(\omega_j\tau-\omega_k\tau)-\frac{1}{2}(S_x^j S_z^k-S_z^j S_x^k)\sin(\omega_j\tau-\omega_k\tau)]/|k-j|^3+\frac{3}{4}\sum_{j<k}[-(S_x^j S_z^k+S_z^j S_x^k)\cos(\omega_j\tau+\omega_k\tau)+(S_z^j S_z^k-S_x^j S_x^k)\sin(\omega_j\tau+\omega_k\tau)]/|k-j|^3$.
When the transport Hamiltonian does not contain a disorder field, only the term $\frac{3}{4}\sum_{j<k}(S_z^j S_z^k-S_x^j S_x^k)\sin(\omega_j\tau+\omega_k\tau)/|k-j|^3$ in $\mathcal{O}_{7'}$ has nonzero overlap with the initial state, therefore $\mathcal{O}=\frac{3}{4}\sum_{j<k}(S_z^j S_z^k-S_x^j S_x^k)\sin(\omega_j\tau+\omega_k\tau)/|k-j|^3\approx\frac{3}{4}\sum_{j}(S_z^j S_z^{j+1}-S_x^j S_x^{j+1})\sin(\omega_j\tau+\omega_k\tau)$.
\item[$\bullet$] Signal: The final signal after phase cycling is $S=\mathrm{Tr}(\rho(t)\mathcal{O})$, with both $\rho$ and $\mathcal{O}$ of the form $rDQ_y$. The random coefficient $\alpha_{j}'=\sin(\omega_j \tau+\omega_k\tau)$.
\item[$\bullet$] Variations: Random double quantum state in xy place can be engineered with rotation of all pulses (except the transport Hamiltonian engineering pulses) along z axis. Autocorrelation of random double quantum state along z axis can be measure by moving the transport step after the $\pi/2$ pulse between step $7'$ and $6'$.
\end{enumerate}
\section{Properties of random states}
\label{sec:properties}
In this section, we analyze various properties of the random state both theoretically and experimentally. The agreement of theory and experiment also serves as a verification that we indeed create the state we expect.
\subsection{Random state magnitude vs preparation cycles}
\label{sec:MagNprep}
\begin{figure}[!htp]
    \centering
    \includegraphics[width=0.47\textwidth]{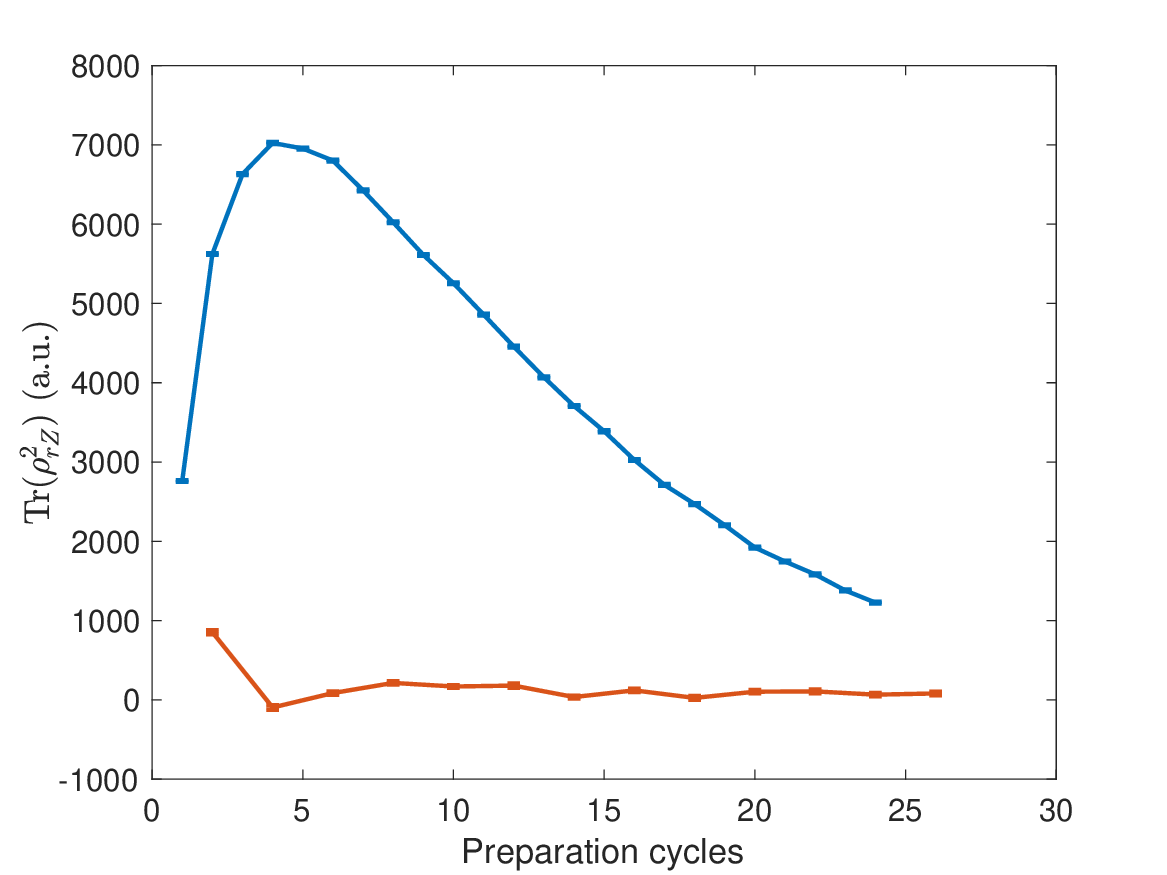}
    \includegraphics[width=0.47\textwidth]{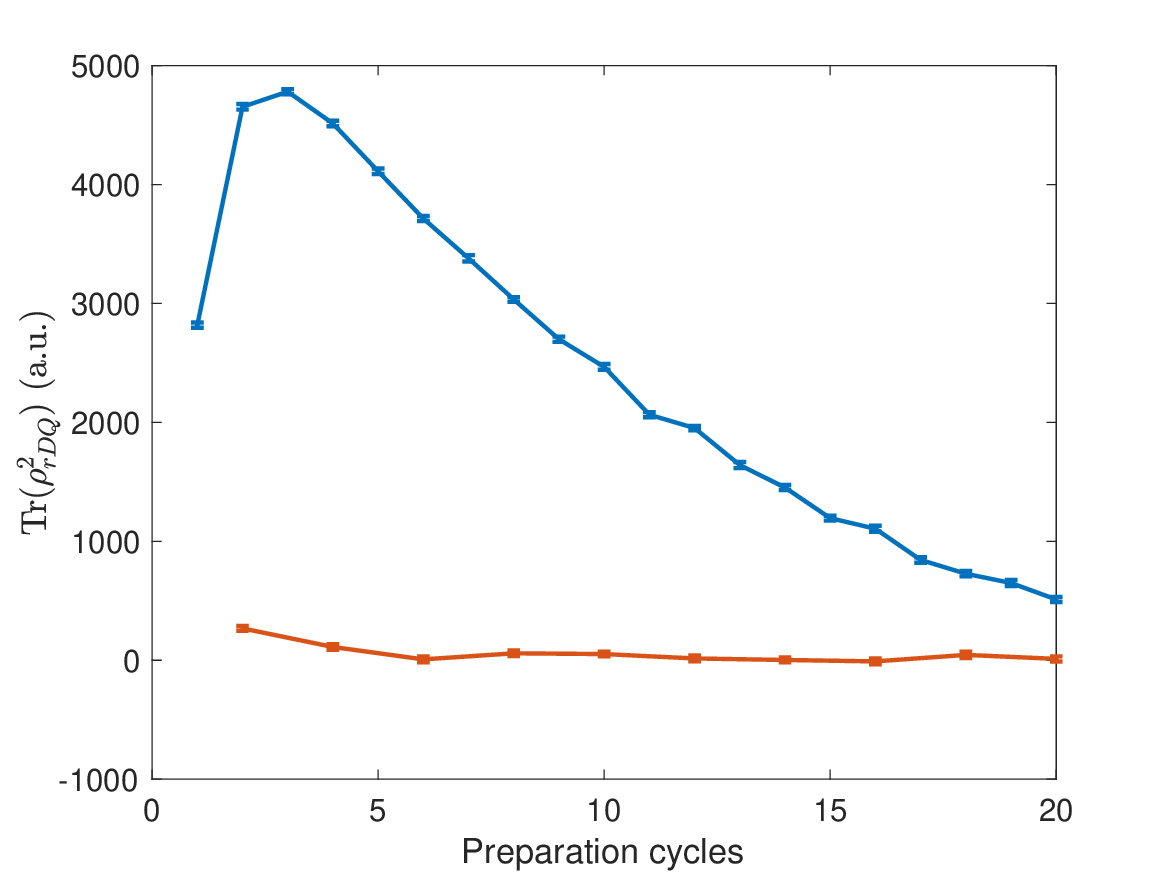}
    \caption{Random Zeeman (a) and double quantum (b) state magnitude as a function of preparation cycles. Blue curves are obtained from experiments that both initial state and observable are engineered with regular WAHUHA8 sequence, while initial states for red curves are engineered with additional $\pi$ pulses after every WAHUHA8 cycles.}
    \label{fig:wZwZ_Nprep_pi}
\end{figure}
Here we study the random state magnitude as a function of preparation cycles. The number of WAHUHA8 cycles used to refocus random states is the same as the preparation cycles. Blue curves in Fig.~\ref{fig:wZwZ_Nprep_pi} show that for both random Zeeman state and random double quantum state, there is an initial raise of magnitude followed by a slow decrease. The raise signals the transformation from initial homogeneous state to random state, while the decrease is a result of imperfection of the sequence. As a control, we add an additional $\pi$ pulse along y axis after every WAHUHA8 cycle during the initial state preparation~\footnote{The $\pi$ pulse is implemented by changing the last $\pi$/2 pulse of WAHUHA8 from -y to y. Compared to physically applying a $\pi$ pulse, the phase change does not elongate the sequence thus is more robust.}, while keeping the observable engineering part unchanged. The $\pi$ pulses refocus the disorder field Hamiltonian during the state preparation at even cycles and thus no random state is created. As a result, the measured random state magnitude is close to zero as shown by the red curves in Fig.~\ref{fig:wZwZ_Nprep_pi}.

\subsection{Multiple quantum coherence}
A general Hermitian observable can be expanded in Pauli string basis
\begin{equation}
    \mathcal{O}_\rho=\sum_s a_s \sigma_s,
\end{equation}
where $s$ is a tuple $s=(\alpha_1,\alpha_2,\cdots,\alpha_L)$ and $\alpha_j=0,x,y,z$; $a_s$ is a scalar coefficient (may be random); $\sigma_s=\otimes_{j=1}^{L}\sigma_{\alpha_j}$ is Pauli string with $\sigma_0=\mathbb{I}$. Here we use Pauli operators $\sigma_\alpha^j=2S_\alpha^j$ so that the normalization factors is a constant for all $s$, $\mathrm{Tr}(\sigma_s^2)=2^L$. 
As we are interested in autocorrelation functions, it is sufficient to evaluate the correlation $C_{ss'}=\mathbb{E}(a_s a_{s'})$. In particular, by diagonalizing the correlation matrix $C_{ss'}$, we can get the principal components of the random state
\begin{equation}
    \mathcal{O}_{\rho}=\sum_\mu d_\mu \mathcal{T}_\mu,
\end{equation}
where $d_\mu$ are independent random variables $\mathbb{E}(d_\mu d_\nu)=\Lambda_\mu \delta_{\mu\nu}$, with $\Lambda_\mu$ being the eigenvalues of $C_{ss'}$ in decreasing order; $\mathcal{T}_\mu$ are orthonormal operators that we name as principal components, $\mathrm{Tr}(\mathcal{T}_\mu\mathcal{T}_\nu^\dagger)=2^L \delta_{\mu\nu}$.
As we do not have universal control, we cannot run a state tomography to determine $C_{ss'}$. Instead, we diagnose the state by global rotations, as different Pauli strings respond differently to global rotations. With some reasonable assumptions, we can then derive $C_{ss'}$.

Global rotations are commonly used in solid-state NMR to analyze spin states, a technique known as multiple quantum coherence (MQC)~\cite{Suter87,Cho03}. MQC characterizes the state by its response to global rotations, therefore it is useful to write the state in the basis of irreducible spherical tensor operators (ISTOs) instead of Pauli strings
\begin{equation}\label{eq:WignerD}
    \mathcal{O}_\rho=\sum_{lm\lambda} a^{(\lambda)}_{lm}T^{(\lambda)}_{lm},
\end{equation}
where $l=1,2,\cdots$, $m=-l,-l+1,\cdots,l$,  $a^{(\lambda)}_{lm}$ are scalar coefficients. $\lambda$ is an additional index that labels different ISTOs in the spin chain with the same values $l$ and $m$. The ISTOs form an orthonormal basis, $\mathrm{Tr}[T^{(\lambda)}_{lm}(T^{(\lambda')}_{l'm'})^\dagger]=2^L\delta_{\lambda\lambda'}\delta_{ll'}\delta_{mm'}$. The ISTOs are defined by the following rotational property
\begin{equation}
    U(\phi,\theta,\gamma)T^{(\lambda)}_{lm}U^\dagger(\phi,\theta,\gamma)=\sum_{m'}D_{lmm'}(\phi,\theta,\gamma)T^{(\lambda)}_{lm},
\end{equation}
where $U(\phi,\theta,\gamma)=e^{-i\phi\sum_jS_z^j}e^{-i\theta\sum_jS_y^j}e^{-i\gamma\sum_jS_z^j}$, $D_{lmm'}(\phi,\theta,\gamma)=e^{-im'\phi}d_{lmm'}(\theta)e^{-im\gamma}$ is the Wigner D-matrix~\cite{Wigner12}. We list ISTOs for the $j^{th}$ nearest-neighbor spin pair,  with  $\lambda=j=1,2,\cdots,L$
\begin{equation}\label{eq:ISTOlist}
    \begin{aligned}
        T^{(j)}_{00}&=\mathbb{I}\\
        T^{(j)}_{10}&=\sigma_z^j\\
        T^{(j)}_{11}&=\sigma_+^j\\
        T^{(j)}_{20}&=\frac{1}{\sqrt{6}}(2\sigma_z^j\sigma_z^{j+1}-\sigma_x^j\sigma_x^{j+1}-\sigma_y^j\sigma_y^{j+1})\\
        T^{(j)}_{21}&=\frac{1}{\sqrt{2}}(\sigma_z^j\sigma_+^{j+1}+\sigma_+^j\sigma_z^{j+1})\\
        T^{(j)}_{22}&=\sigma_+^j \sigma_+^{j+1}.
    \end{aligned}
\end{equation}
ISTOs with $m<0$ can be obtained via $Y_{l,-m}=(-1)^m Y_{lm}^\dagger$. We consider periodic boundary condition here. These ISTOs have the shortest correlation length (distance between furthest non-trivial Pauli operators).
ISTOs with longer correlation length and same $l,m$ can be formed by (i) multiplying by  $(\sigma_x^k\sigma_x^l+\sigma_y^k\sigma_y^l+\sigma_z^k\sigma_z^l)$ (ii) inserting identities between nontrivial Pauli operators. Similarly, ISTO with larger $l,m$ but shortest correlation lengths can be written in terms of triplets, and multiplets of contiguous spins. 

In experiments, we rotate the state $\mathcal{O}_\rho$ along the z axis by an angle $\gamma$, then along y axis by an angle $\theta$, then along y axis by $\phi$ and finally we measure the overlap of the rotated density matrix with the original one. The signal is
\begin{equation}\label{eq:MQCsignal}
    I(\phi,\theta,\gamma)=\mathrm{Tr}\left[U(\phi,\theta,\gamma)\mathcal{O}_\rho U^\dagger(\phi,\theta,\gamma)\mathcal{O}_\rho\right].
\end{equation}

From experiments involving global rotation only, such as MQC experiment, one can distinguish between ISTOs with different $l$ and/or $m$~\cite{vanBeek05}, but it is fundamentally impossible to distinguish those with same $l$ and $m$ but different $\lambda$. Therefore we make the following assumptions. First, for given $l$ and $m$, we consider only ISTOs with the smallest correlation length. Indeed, during the random-state preparation  only rotations and disorder field are applied, which do not create many-body correlation. Operators with longer correlation length can be created only due to higher order effects of the Floquet engineering sequence or experimental imperfections. Therefore, if we detect a spherical component with a given $l,m$, it is most likely to be from the ISTOs with shortest correlation length. With the above assumption, we can consider only $\lambda=1,2,\cdots,L$,
\begin{equation}\label{eq:rhosimp}
    \mathcal{O}_\rho\approx\sum_{j=1}^L\sum_{l=0}^{\infty}\sum_{m=-l}^{l} a_{lm}^{(j)}T_{lm}^{(j)}.
\end{equation}
For given $l$ and $m$, the terms $T_{lm}^{(j)}$ are related by spatial translation. The second assumption is that the state is statistically translational invariant, meaning although the coefficient $a_{lm}^{(j)}$ depends on $j$, its statistics is independent of $j$. This is reasonable because we use a macroscopic crystal, and the chains are only interrupted by rare defects. For our purpose, we only use the translation invariance of correlation, $\mathbb{E}(a_{lm}^{(j)}a_{lm'}^{(j)})=c_{lmm'}$. Using Eq.~\ref{eq:ISTOlist} we can then get $C_{ss'}$ from $c_{lmm'}$.

Next we discuss how to extract $c_{lmm'}$ from experimental signal $I(\phi,\theta,\gamma)$.
Plugging the simplified density matrix in Eq.~\ref{eq:rhosimp} into the signal in Eq.~\ref{eq:MQCsignal}, we obtain 
\begin{equation}
    I(\phi,\theta,\gamma)=L\sum_{l=0}^\infty \sum_{m=-l}^l\sum_{m'=-l}^l c_{lmm'}D_{lmm'}(\phi,\theta,\gamma),
\end{equation}
where we used the fact that $\mathcal{O}_\rho$ is hermitian. 
To extract $c_{lmm'}$, we Fourier transform over $\phi$ and $\gamma$, 
\begin{equation}
    I_{mm'}(\theta)=L\sum_{l=0}^\infty  c_{lmm'}d_{lmm'}(\theta),
\end{equation}
where $I_{mm'}(\theta)=\int_0^{2\pi}d\phi\int_0^{2\pi}d\gamma e^{im'\phi}I(\phi,\theta,\gamma)e^{im\gamma}/(2\pi)^2$. $c_{lmm'}$ can be obtained by fitting to the experimentally measure $I_{mm'}^\mathrm{exp}(\theta)$,
\begin{equation}\label{eq:clmmp}
    c_{lmm'}=\underset{c_{lmm'}}{\operatorname{\mathrm{arg\ min}}}\sum_\theta\left( S_{mm'}^\mathrm{exp}(\theta)-L\sum_{l=0}^\infty  c_{lmm'}d_{lmm'}(\theta)\right)^2.
\end{equation}
$c_{lmm'}=\mathbb{E}(a_{lm}^{(j)}a_{lm'}^{(j)})$ form a correlation matrix, which should be positive semi-definite. However, due to experimental errors, the fitted $c_{lmm'}$ may not be positive semi-definite so we replace the negative eigenvalues by zeros.

\begin{figure}[!htp]
    \centering
    \includegraphics[width=0.47\textwidth]{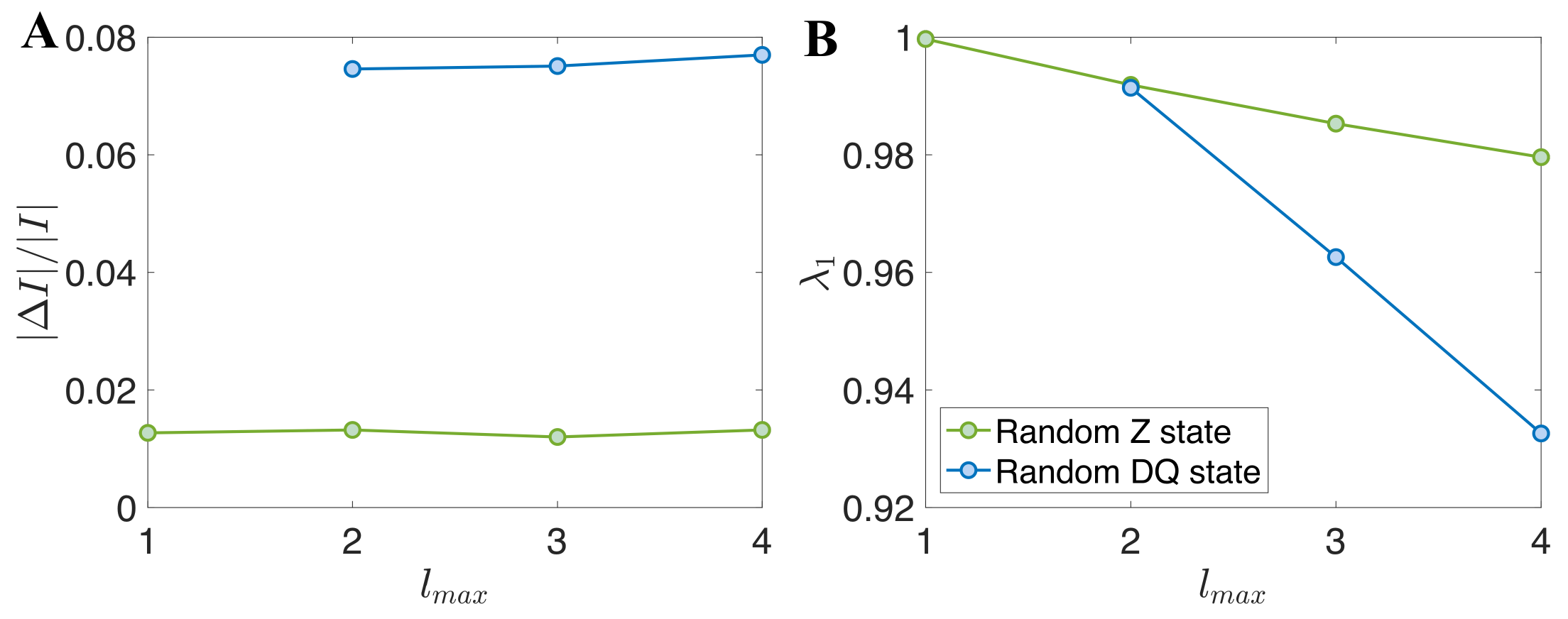}
    \caption{(\textbf{A}) Relative difference the measured $I(\phi,\theta,\gamma)$ and fitted $\hat{I}(\phi,\theta,\gamma)$ as a function of $l_{max}$. $\Delta I=I-\hat{I}$ and $|\cdot|$ denotes Frobenius norm. (\textbf{B}) Largest eigenvalue of the correlation matrix as a function of $l_{max}$. The correlation matrix are normalized such that the positive eigenvalues sum to 1.}
    \label{fig:DsLambda1}
\end{figure}

In experiments, we vary $\phi,\theta,\gamma$ independently from 0 to 345 degree with a step of 45 degree, so we can evaluate $m,m'$ from -3 to 4. To evaluate Eq.~\ref{eq:clmmp}, we assume the maximum correlation length in the prepared states is $l_{max}$, i.e. $a_{lm}=0$ for $l>l_{max}$. In Fig.~\ref{fig:DsLambda1}A, we show the residual of the fitting to $I(\phi,\theta,\gamma)$ as a function of $l_{max}$. For the random Zeeman state (random DQ state), the residual stays stable for $l_{max}\ge 1$ ($l_{max}\ge 2$). Therefore in Fig. 2 in the main text we use $l_{max}=3$ for both states. In Fig.~\ref{fig:DsLambda1}B we show the largest eigenvalue of correlation matrix $C_{SS'}$ as a function of $l_{max}$ and confirm that the largest principal component remains dominant regardless of $l_{max}$.

Finally, we note that for conventional states without randomness, fixing $\gamma=0$ and varying $\phi$ and $\theta$ is sufficient to determine the coefficients $a_{lm}^{(j)}$~\cite{Suter87,Cho03}. The reason is that $a_{lm}^{(j)}$ are deterministic thus $c_{lmm'}$ are dependent, $c_{lmm'}=\sqrt{c_{lmm}c_{lmm'}}$. However, the above equation does not hold for random states and $c_{lmm'}$ contains more degrees of freedom, so independently varying $\phi,\theta$ and $\gamma$ is required here.

\subsection{Spatial correlation}
\label{sec:spatialCorr}
In order to measure local operator correlations, we need the random state coefficients to have zero correlation on different sites. For random Zeeman state $\rho_{rZ}=\sum_j \alpha_j S_z^j$, $\mathbb{E}(\alpha_j \alpha_k)\propto \delta_{jk}$; for random double quantum state $\rho_{rDQ_y}=\sum_{j<k}\alpha_j'(S_z^j S_z^{j+1}-S_x^j S_x^{j+1})$, $\mathbb{E}(\alpha_{j}'\alpha_{k}')\propto \delta_{jk}$. 

We take the random Zeeman state as an example, and the analysis also applies to the random double quantum state. In Sec.~\ref{sec:DetailSeq} we showed that our prepartion protocol sets $\alpha_j=\sin(\omega_j \tau)$, with $\omega_j=(1/3)w_j=(1/3)\sum_{\kappa}J_{j,\kappa}^{FP} I_z^\kappa$. Then, the correlation is
\begin{widetext}
\begin{equation}
\label{eq:SpatialCorr}
\begin{aligned}
    \mathbb{E}(\alpha_j\alpha_k)&=\mathbb{E}[\sin(\omega_j \tau)\sin(\omega_k \tau)]\\
    &=-\frac{1}{4}\mathrm{Tr}\left[\left(\otimes_\kappa e^{i(1/3)J_{j,\kappa}^{FP}I_z^\kappa \tau}-\otimes_\kappa e^{-i(1/3)J_{j,\kappa}^{FP}I_z^\kappa \tau}\right)\left(\otimes_\lambda e^{i(1/3)J_{k,\lambda}^{FP}I_z^\lambda \tau}-\otimes_\lambda e^{-i(1/3)J_{k,\lambda}^{FP}I_z^\lambda \tau}\right)\right]/\mathrm{Tr}(\mathbb{I})\\
    &=-\frac{1}{4}\mathrm{Tr}\left(\otimes_\kappa e^{i(1/3)J_{j,\kappa}^{FP}I_z^\kappa \tau}e^{i(1/3)J_{k,\kappa}^{FP}I_z^\kappa \tau}-\otimes_\kappa e^{-i(1/3)J_{j,\kappa}^{FP}I_z^\kappa \tau}e^{i(1/3)J_{k,\kappa}^{FP}I_z^\kappa \tau}+h.c.\right)/\mathrm{Tr}(\mathbb{I})\\
    &=-\frac{1}{8}\prod_{\kappa}\mathrm{Tr}\left( e^{i(1/3)J_{j,\kappa}^{FP}I_z^\kappa \tau}e^{i(1/3)J_{k,\kappa}^{FP}I_z^\kappa \tau}- e^{-i(1/3)J_{j,\kappa}^{FP}I_z^\kappa \tau}e^{i(1/3)J_{k,\kappa}^{FP}I_z^\kappa \tau}+h.c.\right)\\
    &=\frac{1}{2}\prod_{\kappa} \cos\left[(J_{j,\kappa}^{FP}-J_{k,\kappa}^{FP})\frac{\tau}{6}\right]-\frac{1}{2}\prod_{\kappa}\cos\left[(J_{j,\kappa}^{FP}+J_{k,\kappa}^{FP})\frac{\tau}{6}\right],
\end{aligned}
\end{equation}
\end{widetext}
where we used the fact that $^{31}$P spins are at almost infinite temperature.
The results are most easily understood in two limits. In the $\tau\to0$ limit, the spatial correlation of the random state coefficient equals the spatial correlation of the disordered field $\mathbb{E}(\alpha_j\alpha_k)={\omega_j\omega_k}\tau^2$, which is $\approx0.2$ as shown in Fig.~\ref{fig:disorderfield}B. In the $\tau\to\infty$ limit, the first term of Eq.~\ref{eq:SpatialCorr} is zero for any $j\neq k$, because the large $\tau$ magnifies any tiny difference between $J_{j,\kappa}^{FP}$ and $J_{k,\kappa}^{FP}$ such that  $\left|\cos[(J_{j,\kappa}^{FP}-J_{k,\kappa}^{FP})\frac{\tau}{6}]\right|<1$ for every $\kappa$ and thus the product of all $\kappa$ vanishes. Similarly, we see the second term in Eq.~\ref{eq:SpatialCorr} vanishes for any $j,k$. Therefore, for $\tau\to\infty$, $\mathbb{E}(\alpha_j\alpha_k)=\frac{1}{2}\delta_{jk}$. The above argument can be extend to finite $\tau$: $\mathbb{E}(\alpha_j\alpha_k)=0$ when there are a large number of $\kappa$ satisfying $J_{j,\kappa}^{FP}-J_{k,\kappa}^{FP}>O(1/\tau)$. As a result, the random state grows local for larger $\tau$. Figure~\ref{fig:disorderfield}B shows the example of $k=j+1$. However, larger $\tau$ also leads to smaller random state magnitude (see Sec.~\ref{sec:MagNprep}) and thus a worse signal-to-noise ratio. In practice, we choose the optimal $\tau$ by measuring the dynamics of autocorrelation of random state with different $\tau$ under Hamiltonian with emergent hydrodynamics. The dynamics converges for large $\tau$ and we set $\tau$ at the beginning of the convergence region -- 18 preparation cycles (1.08~ms) for random Zeeman state and 16 cycles (0.96~ms) for random double quantum state.

\section{Data for transport with disorder}
\begin{figure}[!htp]
    \centering
    \includegraphics[width=0.47\textwidth]{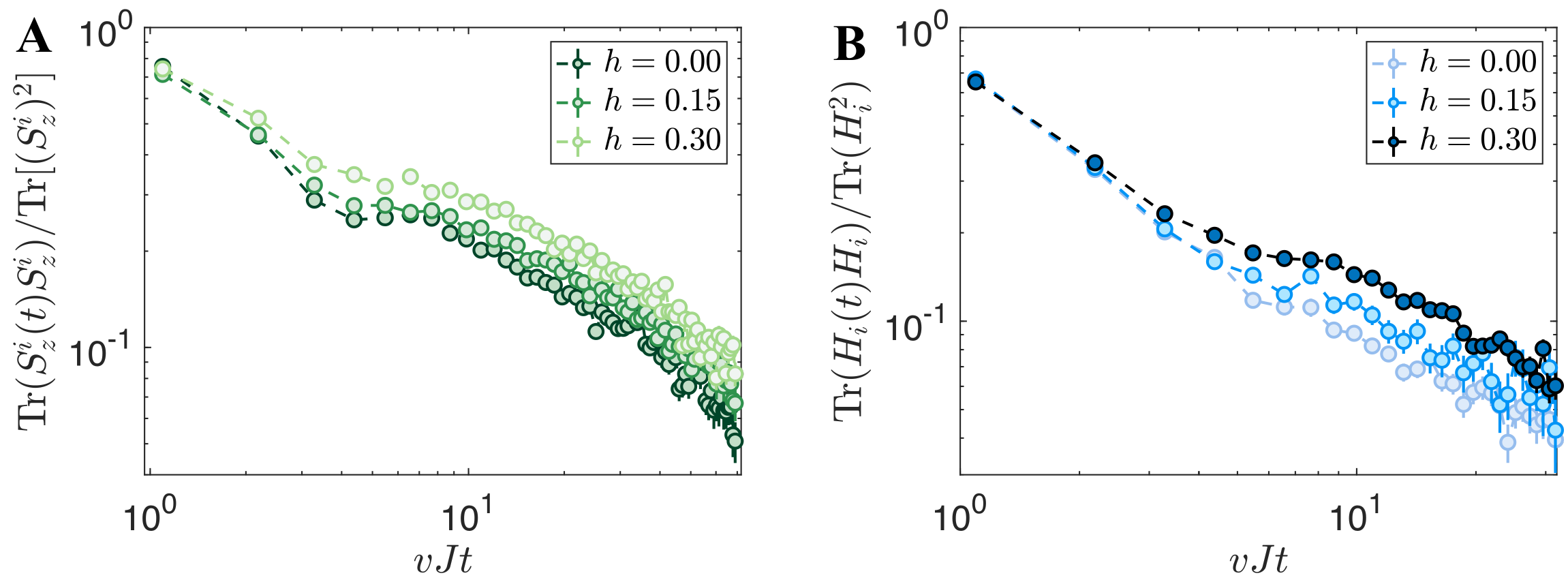}
    \caption{Spin (\textbf{A}) and energy (\textbf{B}) autocorrelation for various disorder field strength $h$.}
    \label{fig:wZwH_h}
\end{figure}

In main text Fig. 4 we show the dynamical exponent for various disorder field strength. We present the source data --- autocorrelation as a function of time here in Fig.~\ref{fig:wZwH_h}.
\bibliography{reference.bib}

\begin{thebibliography}{88}%
\makeatletter
\providecommand \@ifxundefined [1]{%
 \@ifx{#1\undefined}
}%
\providecommand \@ifnum [1]{%
 \ifnum #1\expandafter \@firstoftwo
 \else \expandafter \@secondoftwo
 \fi
}%
\providecommand \@ifx [1]{%
 \ifx #1\expandafter \@firstoftwo
 \else \expandafter \@secondoftwo
 \fi
}%
\providecommand \natexlab [1]{#1}%
\providecommand \enquote  [1]{``#1''}%
\providecommand \bibnamefont  [1]{#1}%
\providecommand \bibfnamefont [1]{#1}%
\providecommand \citenamefont [1]{#1}%
\providecommand \href@noop [0]{\@secondoftwo}%
\providecommand \href [0]{\begingroup \@sanitize@url \@href}%
\providecommand \@href[1]{\@@startlink{#1}\@@href}%
\providecommand \@@href[1]{\endgroup#1\@@endlink}%
\providecommand \@sanitize@url [0]{\catcode `\\12\catcode `\$12\catcode
  `\&12\catcode `\#12\catcode `\^12\catcode `\_12\catcode `\%12\relax}%
\providecommand \@@startlink[1]{}%
\providecommand \@@endlink[0]{}%
\providecommand \url  [0]{\begingroup\@sanitize@url \@url }%
\providecommand \@url [1]{\endgroup\@href {#1}{\urlprefix }}%
\providecommand \urlprefix  [0]{URL }%
\providecommand \Eprint [0]{\href }%
\providecommand \doibase [0]{https://doi.org/}%
\providecommand \selectlanguage [0]{\@gobble}%
\providecommand \bibinfo  [0]{\@secondoftwo}%
\providecommand \bibfield  [0]{\@secondoftwo}%
\providecommand \translation [1]{[#1]}%
\providecommand \BibitemOpen [0]{}%
\providecommand \bibitemStop [0]{}%
\providecommand \bibitemNoStop [0]{.\EOS\space}%
\providecommand \EOS [0]{\spacefactor3000\relax}%
\providecommand \BibitemShut  [1]{\csname bibitem#1\endcsname}%
\let\auto@bib@innerbib\@empty
\bibitem [{\citenamefont {Halliwell}(1999)}]{halliwell1999decoherent}%
  \BibitemOpen
  \bibfield  {author} {\bibinfo {author} {\bibfnamefont {J.}~\bibnamefont
  {Halliwell}},\ }\bibfield  {title} {\bibinfo {title} {Decoherent histories
  and the emergent classicality of local densities},\ }\href@noop {} {\bibfield
   {journal} {\bibinfo  {journal} {Physical review letters}\ }\textbf {\bibinfo
  {volume} {83}},\ \bibinfo {pages} {2481} (\bibinfo {year}
  {1999})}\BibitemShut {NoStop}%
\bibitem [{\citenamefont {Wyatt}(2005)}]{wyatt2005quantum}%
  \BibitemOpen
  \bibfield  {author} {\bibinfo {author} {\bibfnamefont {R.~E.}\ \bibnamefont
  {Wyatt}},\ }\href@noop {} {\emph {\bibinfo {title} {Quantum dynamics with
  trajectories: introduction to quantum hydrodynamics}}},\ Vol.~\bibinfo
  {volume} {28}\ (\bibinfo  {publisher} {Springer Science \& Business Media},\
  \bibinfo {year} {2005})\BibitemShut {NoStop}%
\bibitem [{\citenamefont {Hartle}(2011)}]{hartle2011quasiclassical}%
  \BibitemOpen
  \bibfield  {author} {\bibinfo {author} {\bibfnamefont {J.~B.}\ \bibnamefont
  {Hartle}},\ }\bibfield  {title} {\bibinfo {title} {The quasiclassical realms
  of this quantum universe},\ }\href@noop {} {\bibfield  {journal} {\bibinfo
  {journal} {Foundations of physics}\ }\textbf {\bibinfo {volume} {41}},\
  \bibinfo {pages} {982} (\bibinfo {year} {2011})}\BibitemShut {NoStop}%
\bibitem [{\citenamefont {Spohn}(2012)}]{spohn2012large}%
  \BibitemOpen
  \bibfield  {author} {\bibinfo {author} {\bibfnamefont {H.}~\bibnamefont
  {Spohn}},\ }\href@noop {} {\emph {\bibinfo {title} {Large scale dynamics of
  interacting particles}}}\ (\bibinfo  {publisher} {Springer Science \&
  Business Media},\ \bibinfo {year} {2012})\BibitemShut {NoStop}%
\bibitem [{\citenamefont {Birkhoff}(2015)}]{birkhoff2015hydrodynamics}%
  \BibitemOpen
  \bibfield  {author} {\bibinfo {author} {\bibfnamefont {G.}~\bibnamefont
  {Birkhoff}},\ }\bibfield  {title} {\bibinfo {title} {Hydrodynamics},\ }in\
  \href@noop {} {\emph {\bibinfo {booktitle} {Hydrodynamics}}}\ (\bibinfo
  {publisher} {Princeton University Press},\ \bibinfo {year}
  {2015})\BibitemShut {NoStop}%
\bibitem [{\citenamefont {De~Nardis}\ \emph {et~al.}(2018)\citenamefont
  {De~Nardis}, \citenamefont {Bernard},\ and\ \citenamefont
  {Doyon}}]{de2018hydrodynamic}%
  \BibitemOpen
  \bibfield  {author} {\bibinfo {author} {\bibfnamefont {J.}~\bibnamefont
  {De~Nardis}}, \bibinfo {author} {\bibfnamefont {D.}~\bibnamefont {Bernard}},\
  and\ \bibinfo {author} {\bibfnamefont {B.}~\bibnamefont {Doyon}},\ }\bibfield
   {title} {\bibinfo {title} {Hydrodynamic diffusion in integrable systems},\
  }\href@noop {} {\bibfield  {journal} {\bibinfo  {journal} {Physical review
  letters}\ }\textbf {\bibinfo {volume} {121}},\ \bibinfo {pages} {160603}
  (\bibinfo {year} {2018})}\BibitemShut {NoStop}%
\bibitem [{\citenamefont {Andreev}\ \emph {et~al.}(2011)\citenamefont
  {Andreev}, \citenamefont {Kivelson},\ and\ \citenamefont
  {Spivak}}]{andreev2011hydrodynamic}%
  \BibitemOpen
  \bibfield  {author} {\bibinfo {author} {\bibfnamefont {A.}~\bibnamefont
  {Andreev}}, \bibinfo {author} {\bibfnamefont {S.~A.}\ \bibnamefont
  {Kivelson}},\ and\ \bibinfo {author} {\bibfnamefont {B.}~\bibnamefont
  {Spivak}},\ }\bibfield  {title} {\bibinfo {title} {Hydrodynamic description
  of transport in strongly correlated electron systems},\ }\href@noop {}
  {\bibfield  {journal} {\bibinfo  {journal} {Physical Review Letters}\
  }\textbf {\bibinfo {volume} {106}},\ \bibinfo {pages} {256804} (\bibinfo
  {year} {2011})}\BibitemShut {NoStop}%
\bibitem [{\citenamefont {{\v{Z}}nidari{\v{c}}}\ \emph
  {et~al.}(2016)\citenamefont {{\v{Z}}nidari{\v{c}}}, \citenamefont
  {Scardicchio},\ and\ \citenamefont {Varma}}]{vznidarivc2016diffusive}%
  \BibitemOpen
  \bibfield  {author} {\bibinfo {author} {\bibfnamefont {M.}~\bibnamefont
  {{\v{Z}}nidari{\v{c}}}}, \bibinfo {author} {\bibfnamefont {A.}~\bibnamefont
  {Scardicchio}},\ and\ \bibinfo {author} {\bibfnamefont {V.~K.}\ \bibnamefont
  {Varma}},\ }\bibfield  {title} {\bibinfo {title} {Diffusive and subdiffusive
  spin transport in the ergodic phase of a many-body localizable system},\
  }\href@noop {} {\bibfield  {journal} {\bibinfo  {journal} {Physical review
  letters}\ }\textbf {\bibinfo {volume} {117}},\ \bibinfo {pages} {040601}
  (\bibinfo {year} {2016})}\BibitemShut {NoStop}%
\bibitem [{\citenamefont {Bertini}\ \emph {et~al.}(2016)\citenamefont
  {Bertini}, \citenamefont {Collura}, \citenamefont {De~Nardis},\ and\
  \citenamefont {Fagotti}}]{bertini2016transport}%
  \BibitemOpen
  \bibfield  {author} {\bibinfo {author} {\bibfnamefont {B.}~\bibnamefont
  {Bertini}}, \bibinfo {author} {\bibfnamefont {M.}~\bibnamefont {Collura}},
  \bibinfo {author} {\bibfnamefont {J.}~\bibnamefont {De~Nardis}},\ and\
  \bibinfo {author} {\bibfnamefont {M.}~\bibnamefont {Fagotti}},\ }\bibfield
  {title} {\bibinfo {title} {Transport in out-of-equilibrium x x z chains:
  Exact profiles of charges and currents},\ }\href@noop {} {\bibfield
  {journal} {\bibinfo  {journal} {Physical review letters}\ }\textbf {\bibinfo
  {volume} {117}},\ \bibinfo {pages} {207201} (\bibinfo {year}
  {2016})}\BibitemShut {NoStop}%
\bibitem [{\citenamefont {Leviatan}\ \emph {et~al.}(2017)\citenamefont
  {Leviatan}, \citenamefont {Pollmann}, \citenamefont {Bardarson},
  \citenamefont {Huse},\ and\ \citenamefont {Altman}}]{leviatan2017quantum}%
  \BibitemOpen
  \bibfield  {author} {\bibinfo {author} {\bibfnamefont {E.}~\bibnamefont
  {Leviatan}}, \bibinfo {author} {\bibfnamefont {F.}~\bibnamefont {Pollmann}},
  \bibinfo {author} {\bibfnamefont {J.~H.}\ \bibnamefont {Bardarson}}, \bibinfo
  {author} {\bibfnamefont {D.~A.}\ \bibnamefont {Huse}},\ and\ \bibinfo
  {author} {\bibfnamefont {E.}~\bibnamefont {Altman}},\ }\bibfield  {title}
  {\bibinfo {title} {Quantum thermalization dynamics with matrix-product
  states},\ }\href@noop {} {\bibfield  {journal} {\bibinfo  {journal} {arXiv
  preprint arXiv:1702.08894}\ } (\bibinfo {year} {2017})}\BibitemShut {NoStop}%
\bibitem [{\citenamefont {Ye}\ \emph {et~al.}(2020)\citenamefont {Ye},
  \citenamefont {Machado}, \citenamefont {White}, \citenamefont {Mong},\ and\
  \citenamefont {Yao}}]{ye2020emergent}%
  \BibitemOpen
  \bibfield  {author} {\bibinfo {author} {\bibfnamefont {B.}~\bibnamefont
  {Ye}}, \bibinfo {author} {\bibfnamefont {F.}~\bibnamefont {Machado}},
  \bibinfo {author} {\bibfnamefont {C.~D.}\ \bibnamefont {White}}, \bibinfo
  {author} {\bibfnamefont {R.~S.}\ \bibnamefont {Mong}},\ and\ \bibinfo
  {author} {\bibfnamefont {N.~Y.}\ \bibnamefont {Yao}},\ }\bibfield  {title}
  {\bibinfo {title} {Emergent hydrodynamics in nonequilibrium quantum
  systems},\ }\href@noop {} {\bibfield  {journal} {\bibinfo  {journal}
  {Physical Review Letters}\ }\textbf {\bibinfo {volume} {125}},\ \bibinfo
  {pages} {030601} (\bibinfo {year} {2020})}\BibitemShut {NoStop}%
\bibitem [{\citenamefont {Ljubotina}\ \emph {et~al.}(2019)\citenamefont
  {Ljubotina}, \citenamefont {{\v{Z}}nidari{\v{c}}},\ and\ \citenamefont
  {Prosen}}]{ljubotina2019kardar}%
  \BibitemOpen
  \bibfield  {author} {\bibinfo {author} {\bibfnamefont {M.}~\bibnamefont
  {Ljubotina}}, \bibinfo {author} {\bibfnamefont {M.}~\bibnamefont
  {{\v{Z}}nidari{\v{c}}}},\ and\ \bibinfo {author} {\bibfnamefont
  {T.}~\bibnamefont {Prosen}},\ }\bibfield  {title} {\bibinfo {title}
  {Kardar-parisi-zhang physics in the quantum heisenberg magnet},\ }\href@noop
  {} {\bibfield  {journal} {\bibinfo  {journal} {Physical review letters}\
  }\textbf {\bibinfo {volume} {122}},\ \bibinfo {pages} {210602} (\bibinfo
  {year} {2019})}\BibitemShut {NoStop}%
\bibitem [{\citenamefont {Ye}\ \emph {et~al.}(2022)\citenamefont {Ye},
  \citenamefont {Machado}, \citenamefont {Kemp}, \citenamefont {Hutson},\ and\
  \citenamefont {Yao}}]{ye2022universal}%
  \BibitemOpen
  \bibfield  {author} {\bibinfo {author} {\bibfnamefont {B.}~\bibnamefont
  {Ye}}, \bibinfo {author} {\bibfnamefont {F.}~\bibnamefont {Machado}},
  \bibinfo {author} {\bibfnamefont {J.}~\bibnamefont {Kemp}}, \bibinfo {author}
  {\bibfnamefont {R.~B.}\ \bibnamefont {Hutson}},\ and\ \bibinfo {author}
  {\bibfnamefont {N.~Y.}\ \bibnamefont {Yao}},\ }\bibfield  {title} {\bibinfo
  {title} {Universal kardar-parisi-zhang dynamics in integrable quantum
  systems},\ }\href@noop {} {\bibfield  {journal} {\bibinfo  {journal} {arXiv
  preprint arXiv:2205.02853}\ } (\bibinfo {year} {2022})}\BibitemShut {NoStop}%
\bibitem [{\citenamefont {Sommer}\ \emph {et~al.}(2011)\citenamefont {Sommer},
  \citenamefont {Ku}, \citenamefont {Roati},\ and\ \citenamefont
  {Zwierlein}}]{sommer2011universal}%
  \BibitemOpen
  \bibfield  {author} {\bibinfo {author} {\bibfnamefont {A.}~\bibnamefont
  {Sommer}}, \bibinfo {author} {\bibfnamefont {M.}~\bibnamefont {Ku}}, \bibinfo
  {author} {\bibfnamefont {G.}~\bibnamefont {Roati}},\ and\ \bibinfo {author}
  {\bibfnamefont {M.~W.}\ \bibnamefont {Zwierlein}},\ }\bibfield  {title}
  {\bibinfo {title} {Universal spin transport in a strongly interacting fermi
  gas},\ }\href@noop {} {\bibfield  {journal} {\bibinfo  {journal} {Nature}\
  }\textbf {\bibinfo {volume} {472}},\ \bibinfo {pages} {201} (\bibinfo {year}
  {2011})}\BibitemShut {NoStop}%
\bibitem [{\citenamefont {Moll}\ \emph {et~al.}(2016)\citenamefont {Moll},
  \citenamefont {Kushwaha}, \citenamefont {Nandi}, \citenamefont {Schmidt},\
  and\ \citenamefont {Mackenzie}}]{moll2016evidence}%
  \BibitemOpen
  \bibfield  {author} {\bibinfo {author} {\bibfnamefont {P.~J.}\ \bibnamefont
  {Moll}}, \bibinfo {author} {\bibfnamefont {P.}~\bibnamefont {Kushwaha}},
  \bibinfo {author} {\bibfnamefont {N.}~\bibnamefont {Nandi}}, \bibinfo
  {author} {\bibfnamefont {B.}~\bibnamefont {Schmidt}},\ and\ \bibinfo {author}
  {\bibfnamefont {A.~P.}\ \bibnamefont {Mackenzie}},\ }\bibfield  {title}
  {\bibinfo {title} {Evidence for hydrodynamic electron flow in pdcoo2},\
  }\href@noop {} {\bibfield  {journal} {\bibinfo  {journal} {Science}\ }\textbf
  {\bibinfo {volume} {351}},\ \bibinfo {pages} {1061} (\bibinfo {year}
  {2016})}\BibitemShut {NoStop}%
\bibitem [{\citenamefont {Cepellotti}\ \emph {et~al.}(2015)\citenamefont
  {Cepellotti}, \citenamefont {Fugallo}, \citenamefont {Paulatto},
  \citenamefont {Lazzeri}, \citenamefont {Mauri},\ and\ \citenamefont
  {Marzari}}]{cepellotti2015phonon}%
  \BibitemOpen
  \bibfield  {author} {\bibinfo {author} {\bibfnamefont {A.}~\bibnamefont
  {Cepellotti}}, \bibinfo {author} {\bibfnamefont {G.}~\bibnamefont {Fugallo}},
  \bibinfo {author} {\bibfnamefont {L.}~\bibnamefont {Paulatto}}, \bibinfo
  {author} {\bibfnamefont {M.}~\bibnamefont {Lazzeri}}, \bibinfo {author}
  {\bibfnamefont {F.}~\bibnamefont {Mauri}},\ and\ \bibinfo {author}
  {\bibfnamefont {N.}~\bibnamefont {Marzari}},\ }\bibfield  {title} {\bibinfo
  {title} {Phonon hydrodynamics in two-dimensional materials},\ }\href@noop {}
  {\bibfield  {journal} {\bibinfo  {journal} {Nature communications}\ }\textbf
  {\bibinfo {volume} {6}},\ \bibinfo {pages} {1} (\bibinfo {year}
  {2015})}\BibitemShut {NoStop}%
\bibitem [{\citenamefont {Crossno}\ \emph {et~al.}(2016)\citenamefont
  {Crossno}, \citenamefont {Shi}, \citenamefont {Wang}, \citenamefont {Liu},
  \citenamefont {Harzheim}, \citenamefont {Lucas}, \citenamefont {Sachdev},
  \citenamefont {Kim}, \citenamefont {Taniguchi}, \citenamefont {Watanabe}
  \emph {et~al.}}]{crossno2016observation}%
  \BibitemOpen
  \bibfield  {author} {\bibinfo {author} {\bibfnamefont {J.}~\bibnamefont
  {Crossno}}, \bibinfo {author} {\bibfnamefont {J.~K.}\ \bibnamefont {Shi}},
  \bibinfo {author} {\bibfnamefont {K.}~\bibnamefont {Wang}}, \bibinfo {author}
  {\bibfnamefont {X.}~\bibnamefont {Liu}}, \bibinfo {author} {\bibfnamefont
  {A.}~\bibnamefont {Harzheim}}, \bibinfo {author} {\bibfnamefont
  {A.}~\bibnamefont {Lucas}}, \bibinfo {author} {\bibfnamefont
  {S.}~\bibnamefont {Sachdev}}, \bibinfo {author} {\bibfnamefont
  {P.}~\bibnamefont {Kim}}, \bibinfo {author} {\bibfnamefont {T.}~\bibnamefont
  {Taniguchi}}, \bibinfo {author} {\bibfnamefont {K.}~\bibnamefont {Watanabe}},
  \emph {et~al.},\ }\bibfield  {title} {\bibinfo {title} {Observation of the
  dirac fluid and the breakdown of the wiedemann-franz law in graphene},\
  }\href@noop {} {\bibfield  {journal} {\bibinfo  {journal} {Science}\ }\textbf
  {\bibinfo {volume} {351}},\ \bibinfo {pages} {1058} (\bibinfo {year}
  {2016})}\BibitemShut {NoStop}%
\bibitem [{\citenamefont {Agarwal}\ \emph {et~al.}(2015)\citenamefont
  {Agarwal}, \citenamefont {Gopalakrishnan}, \citenamefont {Knap},
  \citenamefont {M{\"u}ller},\ and\ \citenamefont
  {Demler}}]{agarwal2015anomalous}%
  \BibitemOpen
  \bibfield  {author} {\bibinfo {author} {\bibfnamefont {K.}~\bibnamefont
  {Agarwal}}, \bibinfo {author} {\bibfnamefont {S.}~\bibnamefont
  {Gopalakrishnan}}, \bibinfo {author} {\bibfnamefont {M.}~\bibnamefont
  {Knap}}, \bibinfo {author} {\bibfnamefont {M.}~\bibnamefont {M{\"u}ller}},\
  and\ \bibinfo {author} {\bibfnamefont {E.}~\bibnamefont {Demler}},\
  }\bibfield  {title} {\bibinfo {title} {Anomalous diffusion and griffiths
  effects near the many-body localization transition},\ }\href@noop {}
  {\bibfield  {journal} {\bibinfo  {journal} {Physical review letters}\
  }\textbf {\bibinfo {volume} {114}},\ \bibinfo {pages} {160401} (\bibinfo
  {year} {2015})}\BibitemShut {NoStop}%
\bibitem [{\citenamefont {Castro-Alvaredo}\ \emph {et~al.}(2016)\citenamefont
  {Castro-Alvaredo}, \citenamefont {Doyon},\ and\ \citenamefont
  {Yoshimura}}]{castro2016emergent}%
  \BibitemOpen
  \bibfield  {author} {\bibinfo {author} {\bibfnamefont {O.~A.}\ \bibnamefont
  {Castro-Alvaredo}}, \bibinfo {author} {\bibfnamefont {B.}~\bibnamefont
  {Doyon}},\ and\ \bibinfo {author} {\bibfnamefont {T.}~\bibnamefont
  {Yoshimura}},\ }\bibfield  {title} {\bibinfo {title} {Emergent hydrodynamics
  in integrable quantum systems out of equilibrium},\ }\href@noop {} {\bibfield
   {journal} {\bibinfo  {journal} {Physical Review X}\ }\textbf {\bibinfo
  {volume} {6}},\ \bibinfo {pages} {041065} (\bibinfo {year}
  {2016})}\BibitemShut {NoStop}%
\bibitem [{\citenamefont {Bertini}\ \emph {et~al.}(2021)\citenamefont
  {Bertini}, \citenamefont {Heidrich-Meisner}, \citenamefont {Karrasch},
  \citenamefont {Prosen}, \citenamefont {Steinigeweg},\ and\ \citenamefont
  {{\v{Z}}nidari{\v{c}}}}]{bertini2021finite}%
  \BibitemOpen
  \bibfield  {author} {\bibinfo {author} {\bibfnamefont {B.}~\bibnamefont
  {Bertini}}, \bibinfo {author} {\bibfnamefont {F.}~\bibnamefont
  {Heidrich-Meisner}}, \bibinfo {author} {\bibfnamefont {C.}~\bibnamefont
  {Karrasch}}, \bibinfo {author} {\bibfnamefont {T.}~\bibnamefont {Prosen}},
  \bibinfo {author} {\bibfnamefont {R.}~\bibnamefont {Steinigeweg}},\ and\
  \bibinfo {author} {\bibfnamefont {M.}~\bibnamefont {{\v{Z}}nidari{\v{c}}}},\
  }\bibfield  {title} {\bibinfo {title} {Finite-temperature transport in
  one-dimensional quantum lattice models},\ }\href@noop {} {\bibfield
  {journal} {\bibinfo  {journal} {Reviews of Modern Physics}\ }\textbf
  {\bibinfo {volume} {93}},\ \bibinfo {pages} {025003} (\bibinfo {year}
  {2021})}\BibitemShut {NoStop}%
\bibitem [{\citenamefont {Ilievski}\ and\ \citenamefont
  {De~Nardis}(2017)}]{ilievski2017microscopic}%
  \BibitemOpen
  \bibfield  {author} {\bibinfo {author} {\bibfnamefont {E.}~\bibnamefont
  {Ilievski}}\ and\ \bibinfo {author} {\bibfnamefont {J.}~\bibnamefont
  {De~Nardis}},\ }\bibfield  {title} {\bibinfo {title} {Microscopic origin of
  ideal conductivity in integrable quantum models},\ }\href@noop {} {\bibfield
  {journal} {\bibinfo  {journal} {Physical review letters}\ }\textbf {\bibinfo
  {volume} {119}},\ \bibinfo {pages} {020602} (\bibinfo {year}
  {2017})}\BibitemShut {NoStop}%
\bibitem [{\citenamefont {Gopalakrishnan}\ and\ \citenamefont
  {Vasseur}(2019)}]{gopalakrishnan2019kinetic}%
  \BibitemOpen
  \bibfield  {author} {\bibinfo {author} {\bibfnamefont {S.}~\bibnamefont
  {Gopalakrishnan}}\ and\ \bibinfo {author} {\bibfnamefont {R.}~\bibnamefont
  {Vasseur}},\ }\bibfield  {title} {\bibinfo {title} {Kinetic theory of spin
  diffusion and superdiffusion in x x z spin chains},\ }\href@noop {}
  {\bibfield  {journal} {\bibinfo  {journal} {Physical review letters}\
  }\textbf {\bibinfo {volume} {122}},\ \bibinfo {pages} {127202} (\bibinfo
  {year} {2019})}\BibitemShut {NoStop}%
\bibitem [{\citenamefont {De~Nardis}\ \emph {et~al.}(2019)\citenamefont
  {De~Nardis}, \citenamefont {Bernard},\ and\ \citenamefont
  {Doyon}}]{de2019diffusion}%
  \BibitemOpen
  \bibfield  {author} {\bibinfo {author} {\bibfnamefont {J.}~\bibnamefont
  {De~Nardis}}, \bibinfo {author} {\bibfnamefont {D.}~\bibnamefont {Bernard}},\
  and\ \bibinfo {author} {\bibfnamefont {B.}~\bibnamefont {Doyon}},\ }\bibfield
   {title} {\bibinfo {title} {Diffusion in generalized hydrodynamics and
  quasiparticle scattering},\ }\href@noop {} {\bibfield  {journal} {\bibinfo
  {journal} {SciPost Physics}\ }\textbf {\bibinfo {volume} {6}},\ \bibinfo
  {pages} {049} (\bibinfo {year} {2019})}\BibitemShut {NoStop}%
\bibitem [{\citenamefont {Ilievski}\ \emph {et~al.}(2021)\citenamefont
  {Ilievski}, \citenamefont {De~Nardis}, \citenamefont {Gopalakrishnan},
  \citenamefont {Vasseur},\ and\ \citenamefont
  {Ware}}]{ilievski2021superuniversality}%
  \BibitemOpen
  \bibfield  {author} {\bibinfo {author} {\bibfnamefont {E.}~\bibnamefont
  {Ilievski}}, \bibinfo {author} {\bibfnamefont {J.}~\bibnamefont {De~Nardis}},
  \bibinfo {author} {\bibfnamefont {S.}~\bibnamefont {Gopalakrishnan}},
  \bibinfo {author} {\bibfnamefont {R.}~\bibnamefont {Vasseur}},\ and\ \bibinfo
  {author} {\bibfnamefont {B.}~\bibnamefont {Ware}},\ }\bibfield  {title}
  {\bibinfo {title} {Superuniversality of superdiffusion},\ }\href@noop {}
  {\bibfield  {journal} {\bibinfo  {journal} {Physical Review X}\ }\textbf
  {\bibinfo {volume} {11}},\ \bibinfo {pages} {031023} (\bibinfo {year}
  {2021})}\BibitemShut {NoStop}%
\bibitem [{\citenamefont {De~Nardis}\ \emph {et~al.}(2021)\citenamefont
  {De~Nardis}, \citenamefont {Gopalakrishnan}, \citenamefont {Vasseur},\ and\
  \citenamefont {Ware}}]{de2021stability}%
  \BibitemOpen
  \bibfield  {author} {\bibinfo {author} {\bibfnamefont {J.}~\bibnamefont
  {De~Nardis}}, \bibinfo {author} {\bibfnamefont {S.}~\bibnamefont
  {Gopalakrishnan}}, \bibinfo {author} {\bibfnamefont {R.}~\bibnamefont
  {Vasseur}},\ and\ \bibinfo {author} {\bibfnamefont {B.}~\bibnamefont
  {Ware}},\ }\bibfield  {title} {\bibinfo {title} {Stability of superdiffusion
  in nearly integrable spin chains},\ }\href@noop {} {\bibfield  {journal}
  {\bibinfo  {journal} {Physical review letters}\ }\textbf {\bibinfo {volume}
  {127}},\ \bibinfo {pages} {057201} (\bibinfo {year} {2021})}\BibitemShut
  {NoStop}%
\bibitem [{\citenamefont {Friedman}\ \emph {et~al.}(2020)\citenamefont
  {Friedman}, \citenamefont {Gopalakrishnan},\ and\ \citenamefont
  {Vasseur}}]{friedman2020diffusive}%
  \BibitemOpen
  \bibfield  {author} {\bibinfo {author} {\bibfnamefont {A.~J.}\ \bibnamefont
  {Friedman}}, \bibinfo {author} {\bibfnamefont {S.}~\bibnamefont
  {Gopalakrishnan}},\ and\ \bibinfo {author} {\bibfnamefont {R.}~\bibnamefont
  {Vasseur}},\ }\bibfield  {title} {\bibinfo {title} {Diffusive hydrodynamics
  from integrability breaking},\ }\href@noop {} {\bibfield  {journal} {\bibinfo
   {journal} {Physical Review B}\ }\textbf {\bibinfo {volume} {101}},\ \bibinfo
  {pages} {180302} (\bibinfo {year} {2020})}\BibitemShut {NoStop}%
\bibitem [{\citenamefont {Schemmer}\ \emph {et~al.}(2019)\citenamefont
  {Schemmer}, \citenamefont {Bouchoule}, \citenamefont {Doyon},\ and\
  \citenamefont {Dubail}}]{schemmer2019generalized}%
  \BibitemOpen
  \bibfield  {author} {\bibinfo {author} {\bibfnamefont {M.}~\bibnamefont
  {Schemmer}}, \bibinfo {author} {\bibfnamefont {I.}~\bibnamefont {Bouchoule}},
  \bibinfo {author} {\bibfnamefont {B.}~\bibnamefont {Doyon}},\ and\ \bibinfo
  {author} {\bibfnamefont {J.}~\bibnamefont {Dubail}},\ }\bibfield  {title}
  {\bibinfo {title} {Generalized hydrodynamics on an atom chip},\ }\href@noop
  {} {\bibfield  {journal} {\bibinfo  {journal} {Physical review letters}\
  }\textbf {\bibinfo {volume} {122}},\ \bibinfo {pages} {090601} (\bibinfo
  {year} {2019})}\BibitemShut {NoStop}%
\bibitem [{\citenamefont {Zu}\ \emph {et~al.}(2021)\citenamefont {Zu},
  \citenamefont {Machado}, \citenamefont {Ye}, \citenamefont {Choi},
  \citenamefont {Kobrin}, \citenamefont {Mittiga}, \citenamefont {Hsieh},
  \citenamefont {Bhattacharyya}, \citenamefont {Markham}, \citenamefont
  {Twitchen} \emph {et~al.}}]{zu2021emergent}%
  \BibitemOpen
  \bibfield  {author} {\bibinfo {author} {\bibfnamefont {C.}~\bibnamefont
  {Zu}}, \bibinfo {author} {\bibfnamefont {F.}~\bibnamefont {Machado}},
  \bibinfo {author} {\bibfnamefont {B.}~\bibnamefont {Ye}}, \bibinfo {author}
  {\bibfnamefont {S.}~\bibnamefont {Choi}}, \bibinfo {author} {\bibfnamefont
  {B.}~\bibnamefont {Kobrin}}, \bibinfo {author} {\bibfnamefont
  {T.}~\bibnamefont {Mittiga}}, \bibinfo {author} {\bibfnamefont
  {S.}~\bibnamefont {Hsieh}}, \bibinfo {author} {\bibfnamefont
  {P.}~\bibnamefont {Bhattacharyya}}, \bibinfo {author} {\bibfnamefont
  {M.}~\bibnamefont {Markham}}, \bibinfo {author} {\bibfnamefont
  {D.}~\bibnamefont {Twitchen}}, \emph {et~al.},\ }\bibfield  {title} {\bibinfo
  {title} {Emergent hydrodynamics in a strongly interacting dipolar spin
  ensemble},\ }\href@noop {} {\bibfield  {journal} {\bibinfo  {journal}
  {Nature}\ }\textbf {\bibinfo {volume} {597}},\ \bibinfo {pages} {45}
  (\bibinfo {year} {2021})}\BibitemShut {NoStop}%
\bibitem [{\citenamefont {Malvania}\ \emph {et~al.}(2021)\citenamefont
  {Malvania}, \citenamefont {Zhang}, \citenamefont {Le}, \citenamefont
  {Dubail}, \citenamefont {Rigol},\ and\ \citenamefont
  {Weiss}}]{malvania2021generalized}%
  \BibitemOpen
  \bibfield  {author} {\bibinfo {author} {\bibfnamefont {N.}~\bibnamefont
  {Malvania}}, \bibinfo {author} {\bibfnamefont {Y.}~\bibnamefont {Zhang}},
  \bibinfo {author} {\bibfnamefont {Y.}~\bibnamefont {Le}}, \bibinfo {author}
  {\bibfnamefont {J.}~\bibnamefont {Dubail}}, \bibinfo {author} {\bibfnamefont
  {M.}~\bibnamefont {Rigol}},\ and\ \bibinfo {author} {\bibfnamefont {D.~S.}\
  \bibnamefont {Weiss}},\ }\bibfield  {title} {\bibinfo {title} {Generalized
  hydrodynamics in strongly interacting 1d bose gases},\ }\href@noop {}
  {\bibfield  {journal} {\bibinfo  {journal} {Science}\ }\textbf {\bibinfo
  {volume} {373}},\ \bibinfo {pages} {1129} (\bibinfo {year}
  {2021})}\BibitemShut {NoStop}%
\bibitem [{\citenamefont {Wei}\ \emph {et~al.}(2022)\citenamefont {Wei},
  \citenamefont {Rubio-Abadal}, \citenamefont {Ye}, \citenamefont {Machado},
  \citenamefont {Kemp}, \citenamefont {Srakaew}, \citenamefont {Hollerith},
  \citenamefont {Rui}, \citenamefont {Gopalakrishnan}, \citenamefont {Yao}
  \emph {et~al.}}]{wei2022quantum}%
  \BibitemOpen
  \bibfield  {author} {\bibinfo {author} {\bibfnamefont {D.}~\bibnamefont
  {Wei}}, \bibinfo {author} {\bibfnamefont {A.}~\bibnamefont {Rubio-Abadal}},
  \bibinfo {author} {\bibfnamefont {B.}~\bibnamefont {Ye}}, \bibinfo {author}
  {\bibfnamefont {F.}~\bibnamefont {Machado}}, \bibinfo {author} {\bibfnamefont
  {J.}~\bibnamefont {Kemp}}, \bibinfo {author} {\bibfnamefont {K.}~\bibnamefont
  {Srakaew}}, \bibinfo {author} {\bibfnamefont {S.}~\bibnamefont {Hollerith}},
  \bibinfo {author} {\bibfnamefont {J.}~\bibnamefont {Rui}}, \bibinfo {author}
  {\bibfnamefont {S.}~\bibnamefont {Gopalakrishnan}}, \bibinfo {author}
  {\bibfnamefont {N.~Y.}\ \bibnamefont {Yao}}, \emph {et~al.},\ }\bibfield
  {title} {\bibinfo {title} {Quantum gas microscopy of kardar-parisi-zhang
  superdiffusion},\ }\href@noop {} {\bibfield  {journal} {\bibinfo  {journal}
  {Science}\ }\textbf {\bibinfo {volume} {376}},\ \bibinfo {pages} {716}
  (\bibinfo {year} {2022})}\BibitemShut {NoStop}%
\bibitem [{\citenamefont {Joshi}\ \emph {et~al.}(2022)\citenamefont {Joshi},
  \citenamefont {Kranzl}, \citenamefont {Schuckert}, \citenamefont {Lovas},
  \citenamefont {Maier}, \citenamefont {Blatt}, \citenamefont {Knap},\ and\
  \citenamefont {Roos}}]{joshi2022observing}%
  \BibitemOpen
  \bibfield  {author} {\bibinfo {author} {\bibfnamefont {M.~K.}\ \bibnamefont
  {Joshi}}, \bibinfo {author} {\bibfnamefont {F.}~\bibnamefont {Kranzl}},
  \bibinfo {author} {\bibfnamefont {A.}~\bibnamefont {Schuckert}}, \bibinfo
  {author} {\bibfnamefont {I.}~\bibnamefont {Lovas}}, \bibinfo {author}
  {\bibfnamefont {C.}~\bibnamefont {Maier}}, \bibinfo {author} {\bibfnamefont
  {R.}~\bibnamefont {Blatt}}, \bibinfo {author} {\bibfnamefont
  {M.}~\bibnamefont {Knap}},\ and\ \bibinfo {author} {\bibfnamefont {C.~F.}\
  \bibnamefont {Roos}},\ }\bibfield  {title} {\bibinfo {title} {Observing
  emergent hydrodynamics in a long-range quantum magnet},\ }\href@noop {}
  {\bibfield  {journal} {\bibinfo  {journal} {Science}\ }\textbf {\bibinfo
  {volume} {376}},\ \bibinfo {pages} {720} (\bibinfo {year}
  {2022})}\BibitemShut {NoStop}%
\bibitem [{\citenamefont {Altman}\ \emph {et~al.}(2021)\citenamefont {Altman},
  \citenamefont {Brown}, \citenamefont {Carleo}, \citenamefont {Carr},
  \citenamefont {Demler}, \citenamefont {Chin}, \citenamefont {DeMarco},
  \citenamefont {Economou}, \citenamefont {Eriksson}, \citenamefont {Fu},
  \citenamefont {Greiner}, \citenamefont {Hazzard}, \citenamefont {Hulet},
  \citenamefont {Koll\'ar}, \citenamefont {Lev}, \citenamefont {Lukin},
  \citenamefont {Ma}, \citenamefont {Mi}, \citenamefont {Misra}, \citenamefont
  {Monroe}, \citenamefont {Murch}, \citenamefont {Nazario}, \citenamefont {Ni},
  \citenamefont {Potter}, \citenamefont {Roushan}, \citenamefont {Saffman},
  \citenamefont {Schleier-Smith}, \citenamefont {Siddiqi}, \citenamefont
  {Simmonds}, \citenamefont {Singh}, \citenamefont {Spielman}, \citenamefont
  {Temme}, \citenamefont {Weiss}, \citenamefont {Vu\ifmmode \check{c}\else
  \v{c}\fi{}kovi\ifmmode~\acute{c}\else \'{c}\fi{}}, \citenamefont
  {Vuleti\ifmmode~\acute{c}\else \'{c}\fi{}}, \citenamefont {Ye},\ and\
  \citenamefont {Zwierlein}}]{Altman21}%
  \BibitemOpen
  \bibfield  {author} {\bibinfo {author} {\bibfnamefont {E.}~\bibnamefont
  {Altman}}, \bibinfo {author} {\bibfnamefont {K.~R.}\ \bibnamefont {Brown}},
  \bibinfo {author} {\bibfnamefont {G.}~\bibnamefont {Carleo}}, \bibinfo
  {author} {\bibfnamefont {L.~D.}\ \bibnamefont {Carr}}, \bibinfo {author}
  {\bibfnamefont {E.}~\bibnamefont {Demler}}, \bibinfo {author} {\bibfnamefont
  {C.}~\bibnamefont {Chin}}, \bibinfo {author} {\bibfnamefont {B.}~\bibnamefont
  {DeMarco}}, \bibinfo {author} {\bibfnamefont {S.~E.}\ \bibnamefont
  {Economou}}, \bibinfo {author} {\bibfnamefont {M.~A.}\ \bibnamefont
  {Eriksson}}, \bibinfo {author} {\bibfnamefont {K.-M.~C.}\ \bibnamefont {Fu}},
  \bibinfo {author} {\bibfnamefont {M.}~\bibnamefont {Greiner}}, \bibinfo
  {author} {\bibfnamefont {K.~R.}\ \bibnamefont {Hazzard}}, \bibinfo {author}
  {\bibfnamefont {R.~G.}\ \bibnamefont {Hulet}}, \bibinfo {author}
  {\bibfnamefont {A.~J.}\ \bibnamefont {Koll\'ar}}, \bibinfo {author}
  {\bibfnamefont {B.~L.}\ \bibnamefont {Lev}}, \bibinfo {author} {\bibfnamefont
  {M.~D.}\ \bibnamefont {Lukin}}, \bibinfo {author} {\bibfnamefont
  {R.}~\bibnamefont {Ma}}, \bibinfo {author} {\bibfnamefont {X.}~\bibnamefont
  {Mi}}, \bibinfo {author} {\bibfnamefont {S.}~\bibnamefont {Misra}}, \bibinfo
  {author} {\bibfnamefont {C.}~\bibnamefont {Monroe}}, \bibinfo {author}
  {\bibfnamefont {K.}~\bibnamefont {Murch}}, \bibinfo {author} {\bibfnamefont
  {Z.}~\bibnamefont {Nazario}}, \bibinfo {author} {\bibfnamefont {K.-K.}\
  \bibnamefont {Ni}}, \bibinfo {author} {\bibfnamefont {A.~C.}\ \bibnamefont
  {Potter}}, \bibinfo {author} {\bibfnamefont {P.}~\bibnamefont {Roushan}},
  \bibinfo {author} {\bibfnamefont {M.}~\bibnamefont {Saffman}}, \bibinfo
  {author} {\bibfnamefont {M.}~\bibnamefont {Schleier-Smith}}, \bibinfo
  {author} {\bibfnamefont {I.}~\bibnamefont {Siddiqi}}, \bibinfo {author}
  {\bibfnamefont {R.}~\bibnamefont {Simmonds}}, \bibinfo {author}
  {\bibfnamefont {M.}~\bibnamefont {Singh}}, \bibinfo {author} {\bibfnamefont
  {I.}~\bibnamefont {Spielman}}, \bibinfo {author} {\bibfnamefont
  {K.}~\bibnamefont {Temme}}, \bibinfo {author} {\bibfnamefont {D.~S.}\
  \bibnamefont {Weiss}}, \bibinfo {author} {\bibfnamefont {J.}~\bibnamefont
  {Vu\ifmmode \check{c}\else \v{c}\fi{}kovi\ifmmode~\acute{c}\else
  \'{c}\fi{}}}, \bibinfo {author} {\bibfnamefont {V.}~\bibnamefont
  {Vuleti\ifmmode~\acute{c}\else \'{c}\fi{}}}, \bibinfo {author} {\bibfnamefont
  {J.}~\bibnamefont {Ye}},\ and\ \bibinfo {author} {\bibfnamefont
  {M.}~\bibnamefont {Zwierlein}},\ }\bibfield  {title} {\bibinfo {title}
  {Quantum simulators: Architectures and opportunities},\ }\href
  {https://doi.org/10.1103/PRXQuantum.2.017003} {\bibfield  {journal} {\bibinfo
   {journal} {PRX Quantum}\ }\textbf {\bibinfo {volume} {2}},\ \bibinfo {pages}
  {017003} (\bibinfo {year} {2021})}\BibitemShut {NoStop}%
\bibitem [{\citenamefont {Bakr}\ \emph {et~al.}(2009)\citenamefont {Bakr},
  \citenamefont {Gillen}, \citenamefont {Peng}, \citenamefont {Folling},\ and\
  \citenamefont {Greiner}}]{Bakr09}%
  \BibitemOpen
  \bibfield  {author} {\bibinfo {author} {\bibfnamefont {W.~S.}\ \bibnamefont
  {Bakr}}, \bibinfo {author} {\bibfnamefont {J.~I.}\ \bibnamefont {Gillen}},
  \bibinfo {author} {\bibfnamefont {A.}~\bibnamefont {Peng}}, \bibinfo {author}
  {\bibfnamefont {S.}~\bibnamefont {Folling}},\ and\ \bibinfo {author}
  {\bibfnamefont {M.}~\bibnamefont {Greiner}},\ }\bibfield  {title} {\bibinfo
  {title} {A quantum gas microscope for detecting single atoms in a
  hubbard-regime optical lattice},\ }\href
  {https://doi.org/10.1038/nature08482} {\bibfield  {journal} {\bibinfo
  {journal} {Nature}\ }\textbf {\bibinfo {volume} {462}},\ \bibinfo {pages}
  {74} (\bibinfo {year} {2009})}\BibitemShut {NoStop}%
\bibitem [{\citenamefont {Zhang}\ and\ \citenamefont
  {Cory}(1998)}]{zhang1998first}%
  \BibitemOpen
  \bibfield  {author} {\bibinfo {author} {\bibfnamefont {W.}~\bibnamefont
  {Zhang}}\ and\ \bibinfo {author} {\bibfnamefont {D.}~\bibnamefont {Cory}},\
  }\bibfield  {title} {\bibinfo {title} {First direct measurement of the spin
  diffusion rate in a homogenous solid},\ }\href@noop {} {\bibfield  {journal}
  {\bibinfo  {journal} {Physical review letters}\ }\textbf {\bibinfo {volume}
  {80}},\ \bibinfo {pages} {1324} (\bibinfo {year} {1998})}\BibitemShut
  {NoStop}%
\bibitem [{\citenamefont {Rittweger}\ \emph {et~al.}(2009)\citenamefont
  {Rittweger}, \citenamefont {Han}, \citenamefont {Irvine}, \citenamefont
  {Eggeling},\ and\ \citenamefont {Hell}}]{rittweger2009sted}%
  \BibitemOpen
  \bibfield  {author} {\bibinfo {author} {\bibfnamefont {E.}~\bibnamefont
  {Rittweger}}, \bibinfo {author} {\bibfnamefont {K.~Y.}\ \bibnamefont {Han}},
  \bibinfo {author} {\bibfnamefont {S.~E.}\ \bibnamefont {Irvine}}, \bibinfo
  {author} {\bibfnamefont {C.}~\bibnamefont {Eggeling}},\ and\ \bibinfo
  {author} {\bibfnamefont {S.~W.}\ \bibnamefont {Hell}},\ }\bibfield  {title}
  {\bibinfo {title} {Sted microscopy reveals crystal colour centres with
  nanometric resolution},\ }\href@noop {} {\bibfield  {journal} {\bibinfo
  {journal} {Nature Photonics}\ }\textbf {\bibinfo {volume} {3}},\ \bibinfo
  {pages} {144} (\bibinfo {year} {2009})}\BibitemShut {NoStop}%
\bibitem [{\citenamefont {Maurer}\ \emph {et~al.}(2010)\citenamefont {Maurer},
  \citenamefont {Maze}, \citenamefont {Stanwix}, \citenamefont {Jiang},
  \citenamefont {Gorshkov}, \citenamefont {Zibrov}, \citenamefont {Harke},
  \citenamefont {Hodges}, \citenamefont {Zibrov}, \citenamefont {Yacoby} \emph
  {et~al.}}]{maurer2010far}%
  \BibitemOpen
  \bibfield  {author} {\bibinfo {author} {\bibfnamefont {P.}~\bibnamefont
  {Maurer}}, \bibinfo {author} {\bibfnamefont {J.}~\bibnamefont {Maze}},
  \bibinfo {author} {\bibfnamefont {P.}~\bibnamefont {Stanwix}}, \bibinfo
  {author} {\bibfnamefont {L.}~\bibnamefont {Jiang}}, \bibinfo {author}
  {\bibfnamefont {A.~V.}\ \bibnamefont {Gorshkov}}, \bibinfo {author}
  {\bibfnamefont {A.~A.}\ \bibnamefont {Zibrov}}, \bibinfo {author}
  {\bibfnamefont {B.}~\bibnamefont {Harke}}, \bibinfo {author} {\bibfnamefont
  {J.}~\bibnamefont {Hodges}}, \bibinfo {author} {\bibfnamefont {A.~S.}\
  \bibnamefont {Zibrov}}, \bibinfo {author} {\bibfnamefont {A.}~\bibnamefont
  {Yacoby}}, \emph {et~al.},\ }\bibfield  {title} {\bibinfo {title} {Far-field
  optical imaging and manipulation of individual spins with nanoscale
  resolution},\ }\href@noop {} {\bibfield  {journal} {\bibinfo  {journal}
  {Nature Physics}\ }\textbf {\bibinfo {volume} {6}},\ \bibinfo {pages} {912}
  (\bibinfo {year} {2010})}\BibitemShut {NoStop}%
\bibitem [{\citenamefont {Chen}\ \emph {et~al.}(2013)\citenamefont {Chen},
  \citenamefont {Gaathon}, \citenamefont {Trusheim},\ and\ \citenamefont
  {Englund}}]{chen2013wide}%
  \BibitemOpen
  \bibfield  {author} {\bibinfo {author} {\bibfnamefont {E.~H.}\ \bibnamefont
  {Chen}}, \bibinfo {author} {\bibfnamefont {O.}~\bibnamefont {Gaathon}},
  \bibinfo {author} {\bibfnamefont {M.~E.}\ \bibnamefont {Trusheim}},\ and\
  \bibinfo {author} {\bibfnamefont {D.}~\bibnamefont {Englund}},\ }\bibfield
  {title} {\bibinfo {title} {Wide-field multispectral super-resolution imaging
  using spin-dependent fluorescence in nanodiamonds},\ }\href@noop {}
  {\bibfield  {journal} {\bibinfo  {journal} {Nano letters}\ }\textbf {\bibinfo
  {volume} {13}},\ \bibinfo {pages} {2073} (\bibinfo {year}
  {2013})}\BibitemShut {NoStop}%
\bibitem [{\citenamefont {Pfender}\ \emph {et~al.}(2014)\citenamefont
  {Pfender}, \citenamefont {Aslam}, \citenamefont {Waldherr}, \citenamefont
  {Neumann},\ and\ \citenamefont {Wrachtrup}}]{pfender2014single}%
  \BibitemOpen
  \bibfield  {author} {\bibinfo {author} {\bibfnamefont {M.}~\bibnamefont
  {Pfender}}, \bibinfo {author} {\bibfnamefont {N.}~\bibnamefont {Aslam}},
  \bibinfo {author} {\bibfnamefont {G.}~\bibnamefont {Waldherr}}, \bibinfo
  {author} {\bibfnamefont {P.}~\bibnamefont {Neumann}},\ and\ \bibinfo {author}
  {\bibfnamefont {J.}~\bibnamefont {Wrachtrup}},\ }\bibfield  {title} {\bibinfo
  {title} {Single-spin stochastic optical reconstruction microscopy},\
  }\href@noop {} {\bibfield  {journal} {\bibinfo  {journal} {Proceedings of the
  National Academy of Sciences}\ }\textbf {\bibinfo {volume} {111}},\ \bibinfo
  {pages} {14669} (\bibinfo {year} {2014})}\BibitemShut {NoStop}%
\bibitem [{\citenamefont {Arai}\ \emph {et~al.}(2015)\citenamefont {Arai},
  \citenamefont {Belthangady}, \citenamefont {Zhang}, \citenamefont {Bar-Gill},
  \citenamefont {DeVience}, \citenamefont {Cappellaro}, \citenamefont
  {Yacoby},\ and\ \citenamefont {Walsworth}}]{arai2015fourier}%
  \BibitemOpen
  \bibfield  {author} {\bibinfo {author} {\bibfnamefont {K.}~\bibnamefont
  {Arai}}, \bibinfo {author} {\bibfnamefont {C.}~\bibnamefont {Belthangady}},
  \bibinfo {author} {\bibfnamefont {H.}~\bibnamefont {Zhang}}, \bibinfo
  {author} {\bibfnamefont {N.}~\bibnamefont {Bar-Gill}}, \bibinfo {author}
  {\bibfnamefont {S.}~\bibnamefont {DeVience}}, \bibinfo {author}
  {\bibfnamefont {P.}~\bibnamefont {Cappellaro}}, \bibinfo {author}
  {\bibfnamefont {A.}~\bibnamefont {Yacoby}},\ and\ \bibinfo {author}
  {\bibfnamefont {R.~L.}\ \bibnamefont {Walsworth}},\ }\bibfield  {title}
  {\bibinfo {title} {Fourier magnetic imaging with nanoscale resolution and
  compressed sensing speed-up using electronic spins in diamond},\ }\href@noop
  {} {\bibfield  {journal} {\bibinfo  {journal} {Nature nanotechnology}\
  }\textbf {\bibinfo {volume} {10}},\ \bibinfo {pages} {859} (\bibinfo {year}
  {2015})}\BibitemShut {NoStop}%
\bibitem [{SM()}]{SM}%
  \BibitemOpen
  \href@noop {} {}\bibinfo {note} {For additional details see the Supplementary
  Materials.}\BibitemShut {Stop}%
\bibitem [{\citenamefont {Hunt}(1956)}]{hunt1956some}%
  \BibitemOpen
  \bibfield  {author} {\bibinfo {author} {\bibfnamefont {G.~A.}\ \bibnamefont
  {Hunt}},\ }\bibfield  {title} {\bibinfo {title} {Some theorems concerning
  brownian motion},\ }\href@noop {} {\bibfield  {journal} {\bibinfo  {journal}
  {Transactions of the American Mathematical Society}\ }\textbf {\bibinfo
  {volume} {81}},\ \bibinfo {pages} {294} (\bibinfo {year} {1956})}\BibitemShut
  {NoStop}%
\bibitem [{\citenamefont {Waugh}\ \emph {et~al.}(1968)\citenamefont {Waugh},
  \citenamefont {Huber},\ and\ \citenamefont {Haeberlen}}]{waugh1968approach}%
  \BibitemOpen
  \bibfield  {author} {\bibinfo {author} {\bibfnamefont {J.~S.}\ \bibnamefont
  {Waugh}}, \bibinfo {author} {\bibfnamefont {L.~M.}\ \bibnamefont {Huber}},\
  and\ \bibinfo {author} {\bibfnamefont {U.}~\bibnamefont {Haeberlen}},\
  }\bibfield  {title} {\bibinfo {title} {Approach to high-resolution nmr in
  solids},\ }\href@noop {} {\bibfield  {journal} {\bibinfo  {journal} {Physical
  Review Letters}\ }\textbf {\bibinfo {volume} {20}},\ \bibinfo {pages} {180}
  (\bibinfo {year} {1968})}\BibitemShut {NoStop}%
\bibitem [{\citenamefont {Jeener}\ and\ \citenamefont
  {Broekaert}(1967)}]{jeener1967nuclear}%
  \BibitemOpen
  \bibfield  {author} {\bibinfo {author} {\bibfnamefont {J.}~\bibnamefont
  {Jeener}}\ and\ \bibinfo {author} {\bibfnamefont {P.}~\bibnamefont
  {Broekaert}},\ }\bibfield  {title} {\bibinfo {title} {Nuclear magnetic
  resonance in solids: thermodynamic effects of a pair of rf pulses},\
  }\href@noop {} {\bibfield  {journal} {\bibinfo  {journal} {Physical Review}\
  }\textbf {\bibinfo {volume} {157}},\ \bibinfo {pages} {232} (\bibinfo {year}
  {1967})}\BibitemShut {NoStop}%
\bibitem [{Note1()}]{Note1}%
  \BibitemOpen
  \bibinfo {note} {Here we assume nearest-neighbor coupling for representation
  simplicity, but the results also hold with $1/r^3$ long-range coupling \cite
  {SM}.}\BibitemShut {Stop}%
\bibitem [{\citenamefont {Haeberlen}\ and\ \citenamefont
  {Waugh}(1968)}]{haeberlen1968coherent}%
  \BibitemOpen
  \bibfield  {author} {\bibinfo {author} {\bibfnamefont {U.}~\bibnamefont
  {Haeberlen}}\ and\ \bibinfo {author} {\bibfnamefont {J.~S.}\ \bibnamefont
  {Waugh}},\ }\bibfield  {title} {\bibinfo {title} {Coherent averaging effects
  in magnetic resonance},\ }\href@noop {} {\bibfield  {journal} {\bibinfo
  {journal} {Physical Review}\ }\textbf {\bibinfo {volume} {175}},\ \bibinfo
  {pages} {453} (\bibinfo {year} {1968})}\BibitemShut {NoStop}%
\bibitem [{\citenamefont {Peng}\ \emph {et~al.}(2021)\citenamefont {Peng},
  \citenamefont {Huang}, \citenamefont {Yin}, \citenamefont {Joseph},
  \citenamefont {Ramanathan},\ and\ \citenamefont {Cappellaro}}]{peng2021deep}%
  \BibitemOpen
  \bibfield  {author} {\bibinfo {author} {\bibfnamefont {P.}~\bibnamefont
  {Peng}}, \bibinfo {author} {\bibfnamefont {X.}~\bibnamefont {Huang}},
  \bibinfo {author} {\bibfnamefont {C.}~\bibnamefont {Yin}}, \bibinfo {author}
  {\bibfnamefont {L.}~\bibnamefont {Joseph}}, \bibinfo {author} {\bibfnamefont
  {C.}~\bibnamefont {Ramanathan}},\ and\ \bibinfo {author} {\bibfnamefont
  {P.}~\bibnamefont {Cappellaro}},\ }\bibfield  {title} {\bibinfo {title} {Deep
  reinforcement learning for quantum hamiltonian engineering},\ }\href@noop {}
  {\bibfield  {journal} {\bibinfo  {journal} {arXiv preprint arXiv:2102.13161}\
  } (\bibinfo {year} {2021})}\BibitemShut {NoStop}%
\bibitem [{\citenamefont {Grabowski}\ and\ \citenamefont
  {Mathieu}(1995)}]{grabowski1995structure}%
  \BibitemOpen
  \bibfield  {author} {\bibinfo {author} {\bibfnamefont {M.}~\bibnamefont
  {Grabowski}}\ and\ \bibinfo {author} {\bibfnamefont {P.}~\bibnamefont
  {Mathieu}},\ }\bibfield  {title} {\bibinfo {title} {Structure of the
  conservation laws in quantum integrable spin chains with short range
  interactions},\ }\href@noop {} {\bibfield  {journal} {\bibinfo  {journal}
  {Annals of Physics}\ }\textbf {\bibinfo {volume} {243}},\ \bibinfo {pages}
  {299} (\bibinfo {year} {1995})}\BibitemShut {NoStop}%
\bibitem [{\citenamefont {Zotos}\ \emph {et~al.}(1997)\citenamefont {Zotos},
  \citenamefont {Naef},\ and\ \citenamefont {Prelovsek}}]{zotos1997transport}%
  \BibitemOpen
  \bibfield  {author} {\bibinfo {author} {\bibfnamefont {X.}~\bibnamefont
  {Zotos}}, \bibinfo {author} {\bibfnamefont {F.}~\bibnamefont {Naef}},\ and\
  \bibinfo {author} {\bibfnamefont {P.}~\bibnamefont {Prelovsek}},\ }\bibfield
  {title} {\bibinfo {title} {Transport and conservation laws},\ }\href@noop {}
  {\bibfield  {journal} {\bibinfo  {journal} {Physical Review B}\ }\textbf
  {\bibinfo {volume} {55}},\ \bibinfo {pages} {11029} (\bibinfo {year}
  {1997})}\BibitemShut {NoStop}%
\bibitem [{\citenamefont {Kl{\"u}mper}\ and\ \citenamefont
  {Johnston}(2000)}]{klumper2000thermodynamics}%
  \BibitemOpen
  \bibfield  {author} {\bibinfo {author} {\bibfnamefont {A.}~\bibnamefont
  {Kl{\"u}mper}}\ and\ \bibinfo {author} {\bibfnamefont {D.}~\bibnamefont
  {Johnston}},\ }\bibfield  {title} {\bibinfo {title} {Thermodynamics of the
  spin-1/2 antiferromagnetic uniform heisenberg chain},\ }\href@noop {}
  {\bibfield  {journal} {\bibinfo  {journal} {Physical Review Letters}\
  }\textbf {\bibinfo {volume} {84}},\ \bibinfo {pages} {4701} (\bibinfo {year}
  {2000})}\BibitemShut {NoStop}%
\bibitem [{\citenamefont {Sakai}\ and\ \citenamefont
  {Kl{\"u}mper}(2003)}]{sakai2003non}%
  \BibitemOpen
  \bibfield  {author} {\bibinfo {author} {\bibfnamefont {K.}~\bibnamefont
  {Sakai}}\ and\ \bibinfo {author} {\bibfnamefont {A.}~\bibnamefont
  {Kl{\"u}mper}},\ }\bibfield  {title} {\bibinfo {title} {Non-dissipative
  thermal transport in the massive regimes of the xxz chain},\ }\href@noop {}
  {\bibfield  {journal} {\bibinfo  {journal} {Journal of Physics A:
  Mathematical and General}\ }\textbf {\bibinfo {volume} {36}},\ \bibinfo
  {pages} {11617} (\bibinfo {year} {2003})}\BibitemShut {NoStop}%
\bibitem [{\citenamefont {Prosen}\ and\ \citenamefont
  {{\v{Z}}nidari{\v{c}}}(2009)}]{prosen2009matrix}%
  \BibitemOpen
  \bibfield  {author} {\bibinfo {author} {\bibfnamefont {T.}~\bibnamefont
  {Prosen}}\ and\ \bibinfo {author} {\bibfnamefont {M.}~\bibnamefont
  {{\v{Z}}nidari{\v{c}}}},\ }\bibfield  {title} {\bibinfo {title} {Matrix
  product simulations of non-equilibrium steady states of quantum spin
  chains},\ }\href@noop {} {\bibfield  {journal} {\bibinfo  {journal} {Journal
  of Statistical Mechanics: Theory and Experiment}\ }\textbf {\bibinfo {volume}
  {2009}},\ \bibinfo {pages} {P02035} (\bibinfo {year} {2009})}\BibitemShut
  {NoStop}%
\bibitem [{\citenamefont {Steinigeweg}\ and\ \citenamefont
  {Gemmer}(2009)}]{steinigeweg2009density}%
  \BibitemOpen
  \bibfield  {author} {\bibinfo {author} {\bibfnamefont {R.}~\bibnamefont
  {Steinigeweg}}\ and\ \bibinfo {author} {\bibfnamefont {J.}~\bibnamefont
  {Gemmer}},\ }\bibfield  {title} {\bibinfo {title} {Density dynamics in
  translationally invariant spin-1 2 chains at high temperatures: A
  current-autocorrelation approach to finite time and length scales},\
  }\href@noop {} {\bibfield  {journal} {\bibinfo  {journal} {Physical Review
  B}\ }\textbf {\bibinfo {volume} {80}},\ \bibinfo {pages} {184402} (\bibinfo
  {year} {2009})}\BibitemShut {NoStop}%
\bibitem [{\citenamefont {{\v{Z}}nidari{\v{c}}}(2011)}]{vznidarivc2011spin}%
  \BibitemOpen
  \bibfield  {author} {\bibinfo {author} {\bibfnamefont {M.}~\bibnamefont
  {{\v{Z}}nidari{\v{c}}}},\ }\bibfield  {title} {\bibinfo {title} {Spin
  transport in a one-dimensional anisotropic heisenberg model},\ }\href@noop {}
  {\bibfield  {journal} {\bibinfo  {journal} {Physical Review Letters}\
  }\textbf {\bibinfo {volume} {106}},\ \bibinfo {pages} {220601} (\bibinfo
  {year} {2011})}\BibitemShut {NoStop}%
\bibitem [{\citenamefont {Karrasch}\ \emph {et~al.}(2014)\citenamefont
  {Karrasch}, \citenamefont {Moore},\ and\ \citenamefont
  {Heidrich-Meisner}}]{karrasch2014real}%
  \BibitemOpen
  \bibfield  {author} {\bibinfo {author} {\bibfnamefont {C.}~\bibnamefont
  {Karrasch}}, \bibinfo {author} {\bibfnamefont {J.}~\bibnamefont {Moore}},\
  and\ \bibinfo {author} {\bibfnamefont {F.}~\bibnamefont {Heidrich-Meisner}},\
  }\bibfield  {title} {\bibinfo {title} {Real-time and real-space spin and
  energy dynamics in one-dimensional spin-1 2 systems induced by local quantum
  quenches at finite temperatures},\ }\href@noop {} {\bibfield  {journal}
  {\bibinfo  {journal} {Physical Review B}\ }\textbf {\bibinfo {volume} {89}},\
  \bibinfo {pages} {075139} (\bibinfo {year} {2014})}\BibitemShut {NoStop}%
\bibitem [{\citenamefont {Lucioni}\ \emph {et~al.}(2011)\citenamefont
  {Lucioni}, \citenamefont {Deissler}, \citenamefont {Tanzi}, \citenamefont
  {Roati}, \citenamefont {Zaccanti}, \citenamefont {Modugno}, \citenamefont
  {Larcher}, \citenamefont {Dalfovo}, \citenamefont {Inguscio},\ and\
  \citenamefont {Modugno}}]{lucioni2011observation}%
  \BibitemOpen
  \bibfield  {author} {\bibinfo {author} {\bibfnamefont {E.}~\bibnamefont
  {Lucioni}}, \bibinfo {author} {\bibfnamefont {B.}~\bibnamefont {Deissler}},
  \bibinfo {author} {\bibfnamefont {L.}~\bibnamefont {Tanzi}}, \bibinfo
  {author} {\bibfnamefont {G.}~\bibnamefont {Roati}}, \bibinfo {author}
  {\bibfnamefont {M.}~\bibnamefont {Zaccanti}}, \bibinfo {author}
  {\bibfnamefont {M.}~\bibnamefont {Modugno}}, \bibinfo {author} {\bibfnamefont
  {M.}~\bibnamefont {Larcher}}, \bibinfo {author} {\bibfnamefont
  {F.}~\bibnamefont {Dalfovo}}, \bibinfo {author} {\bibfnamefont
  {M.}~\bibnamefont {Inguscio}},\ and\ \bibinfo {author} {\bibfnamefont
  {G.}~\bibnamefont {Modugno}},\ }\bibfield  {title} {\bibinfo {title}
  {Observation of subdiffusion in a disordered interacting system},\
  }\href@noop {} {\bibfield  {journal} {\bibinfo  {journal} {Physical review
  letters}\ }\textbf {\bibinfo {volume} {106}},\ \bibinfo {pages} {230403}
  (\bibinfo {year} {2011})}\BibitemShut {NoStop}%
\bibitem [{\citenamefont {Vosk}\ \emph {et~al.}(2015)\citenamefont {Vosk},
  \citenamefont {Huse},\ and\ \citenamefont {Altman}}]{vosk2015theory}%
  \BibitemOpen
  \bibfield  {author} {\bibinfo {author} {\bibfnamefont {R.}~\bibnamefont
  {Vosk}}, \bibinfo {author} {\bibfnamefont {D.~A.}\ \bibnamefont {Huse}},\
  and\ \bibinfo {author} {\bibfnamefont {E.}~\bibnamefont {Altman}},\
  }\bibfield  {title} {\bibinfo {title} {Theory of the many-body localization
  transition in one-dimensional systems},\ }\href@noop {} {\bibfield  {journal}
  {\bibinfo  {journal} {Physical Review X}\ }\textbf {\bibinfo {volume} {5}},\
  \bibinfo {pages} {031032} (\bibinfo {year} {2015})}\BibitemShut {NoStop}%
\bibitem [{\citenamefont {Potter}\ \emph {et~al.}(2015)\citenamefont {Potter},
  \citenamefont {Vasseur},\ and\ \citenamefont
  {Parameswaran}}]{potter2015universal}%
  \BibitemOpen
  \bibfield  {author} {\bibinfo {author} {\bibfnamefont {A.~C.}\ \bibnamefont
  {Potter}}, \bibinfo {author} {\bibfnamefont {R.}~\bibnamefont {Vasseur}},\
  and\ \bibinfo {author} {\bibfnamefont {S.}~\bibnamefont {Parameswaran}},\
  }\bibfield  {title} {\bibinfo {title} {Universal properties of many-body
  delocalization transitions},\ }\href@noop {} {\bibfield  {journal} {\bibinfo
  {journal} {Physical Review X}\ }\textbf {\bibinfo {volume} {5}},\ \bibinfo
  {pages} {031033} (\bibinfo {year} {2015})}\BibitemShut {NoStop}%
\bibitem [{\citenamefont {Sahay}\ \emph {et~al.}(2021)\citenamefont {Sahay},
  \citenamefont {Machado}, \citenamefont {Ye}, \citenamefont {Laumann},\ and\
  \citenamefont {Yao}}]{sahay2021emergent}%
  \BibitemOpen
  \bibfield  {author} {\bibinfo {author} {\bibfnamefont {R.}~\bibnamefont
  {Sahay}}, \bibinfo {author} {\bibfnamefont {F.}~\bibnamefont {Machado}},
  \bibinfo {author} {\bibfnamefont {B.}~\bibnamefont {Ye}}, \bibinfo {author}
  {\bibfnamefont {C.~R.}\ \bibnamefont {Laumann}},\ and\ \bibinfo {author}
  {\bibfnamefont {N.~Y.}\ \bibnamefont {Yao}},\ }\bibfield  {title} {\bibinfo
  {title} {Emergent ergodicity at the transition between many-body localized
  phases},\ }\href@noop {} {\bibfield  {journal} {\bibinfo  {journal} {Physical
  review letters}\ }\textbf {\bibinfo {volume} {126}},\ \bibinfo {pages}
  {100604} (\bibinfo {year} {2021})}\BibitemShut {NoStop}%
\bibitem [{\citenamefont {Zaburdaev}\ \emph {et~al.}(2015)\citenamefont
  {Zaburdaev}, \citenamefont {Denisov},\ and\ \citenamefont
  {Klafter}}]{zaburdaev2015levy}%
  \BibitemOpen
  \bibfield  {author} {\bibinfo {author} {\bibfnamefont {V.}~\bibnamefont
  {Zaburdaev}}, \bibinfo {author} {\bibfnamefont {S.}~\bibnamefont {Denisov}},\
  and\ \bibinfo {author} {\bibfnamefont {J.}~\bibnamefont {Klafter}},\
  }\bibfield  {title} {\bibinfo {title} {L{\'e}vy walks},\ }\href@noop {}
  {\bibfield  {journal} {\bibinfo  {journal} {Reviews of Modern Physics}\
  }\textbf {\bibinfo {volume} {87}},\ \bibinfo {pages} {483} (\bibinfo {year}
  {2015})}\BibitemShut {NoStop}%
\bibitem [{\citenamefont {Nahum}\ \emph {et~al.}(2018)\citenamefont {Nahum},
  \citenamefont {Vijay},\ and\ \citenamefont {Haah}}]{nahum2018operator}%
  \BibitemOpen
  \bibfield  {author} {\bibinfo {author} {\bibfnamefont {A.}~\bibnamefont
  {Nahum}}, \bibinfo {author} {\bibfnamefont {S.}~\bibnamefont {Vijay}},\ and\
  \bibinfo {author} {\bibfnamefont {J.}~\bibnamefont {Haah}},\ }\bibfield
  {title} {\bibinfo {title} {Operator spreading in random unitary circuits},\
  }\href@noop {} {\bibfield  {journal} {\bibinfo  {journal} {Physical Review
  X}\ }\textbf {\bibinfo {volume} {8}},\ \bibinfo {pages} {021014} (\bibinfo
  {year} {2018})}\BibitemShut {NoStop}%
\bibitem [{\citenamefont {Von~Keyserlingk}\ \emph {et~al.}(2018)\citenamefont
  {Von~Keyserlingk}, \citenamefont {Rakovszky}, \citenamefont {Pollmann},\ and\
  \citenamefont {Sondhi}}]{von2018operator}%
  \BibitemOpen
  \bibfield  {author} {\bibinfo {author} {\bibfnamefont {C.}~\bibnamefont
  {Von~Keyserlingk}}, \bibinfo {author} {\bibfnamefont {T.}~\bibnamefont
  {Rakovszky}}, \bibinfo {author} {\bibfnamefont {F.}~\bibnamefont
  {Pollmann}},\ and\ \bibinfo {author} {\bibfnamefont {S.~L.}\ \bibnamefont
  {Sondhi}},\ }\bibfield  {title} {\bibinfo {title} {Operator hydrodynamics,
  otocs, and entanglement growth in systems without conservation laws},\
  }\href@noop {} {\bibfield  {journal} {\bibinfo  {journal} {Physical Review
  X}\ }\textbf {\bibinfo {volume} {8}},\ \bibinfo {pages} {021013} (\bibinfo
  {year} {2018})}\BibitemShut {NoStop}%
\bibitem [{\citenamefont {Rakovszky}\ \emph {et~al.}(2018)\citenamefont
  {Rakovszky}, \citenamefont {Pollmann},\ and\ \citenamefont
  {Von~Keyserlingk}}]{rakovszky2018diffusive}%
  \BibitemOpen
  \bibfield  {author} {\bibinfo {author} {\bibfnamefont {T.}~\bibnamefont
  {Rakovszky}}, \bibinfo {author} {\bibfnamefont {F.}~\bibnamefont
  {Pollmann}},\ and\ \bibinfo {author} {\bibfnamefont {C.}~\bibnamefont
  {Von~Keyserlingk}},\ }\bibfield  {title} {\bibinfo {title} {Diffusive
  hydrodynamics of out-of-time-ordered correlators with charge conservation},\
  }\href@noop {} {\bibfield  {journal} {\bibinfo  {journal} {Physical Review
  X}\ }\textbf {\bibinfo {volume} {8}},\ \bibinfo {pages} {031058} (\bibinfo
  {year} {2018})}\BibitemShut {NoStop}%
\bibitem [{\citenamefont {Khemani}\ \emph {et~al.}(2018)\citenamefont
  {Khemani}, \citenamefont {Vishwanath},\ and\ \citenamefont
  {Huse}}]{khemani2018operator}%
  \BibitemOpen
  \bibfield  {author} {\bibinfo {author} {\bibfnamefont {V.}~\bibnamefont
  {Khemani}}, \bibinfo {author} {\bibfnamefont {A.}~\bibnamefont
  {Vishwanath}},\ and\ \bibinfo {author} {\bibfnamefont {D.~A.}\ \bibnamefont
  {Huse}},\ }\bibfield  {title} {\bibinfo {title} {Operator spreading and the
  emergence of dissipative hydrodynamics under unitary evolution with
  conservation laws},\ }\href@noop {} {\bibfield  {journal} {\bibinfo
  {journal} {Physical Review X}\ }\textbf {\bibinfo {volume} {8}},\ \bibinfo
  {pages} {031057} (\bibinfo {year} {2018})}\BibitemShut {NoStop}%
\bibitem [{\citenamefont {Xu}\ and\ \citenamefont
  {Swingle}(2020)}]{xu2020accessing}%
  \BibitemOpen
  \bibfield  {author} {\bibinfo {author} {\bibfnamefont {S.}~\bibnamefont
  {Xu}}\ and\ \bibinfo {author} {\bibfnamefont {B.}~\bibnamefont {Swingle}},\
  }\bibfield  {title} {\bibinfo {title} {Accessing scrambling using matrix
  product operators},\ }\href@noop {} {\bibfield  {journal} {\bibinfo
  {journal} {Nature Physics}\ }\textbf {\bibinfo {volume} {16}},\ \bibinfo
  {pages} {199} (\bibinfo {year} {2020})}\BibitemShut {NoStop}%
\bibitem [{\citenamefont {Xu}\ and\ \citenamefont
  {Swingle}(2019)}]{xu2019locality}%
  \BibitemOpen
  \bibfield  {author} {\bibinfo {author} {\bibfnamefont {S.}~\bibnamefont
  {Xu}}\ and\ \bibinfo {author} {\bibfnamefont {B.}~\bibnamefont {Swingle}},\
  }\bibfield  {title} {\bibinfo {title} {Locality, quantum fluctuations, and
  scrambling},\ }\href@noop {} {\bibfield  {journal} {\bibinfo  {journal}
  {Physical Review X}\ }\textbf {\bibinfo {volume} {9}},\ \bibinfo {pages}
  {031048} (\bibinfo {year} {2019})}\BibitemShut {NoStop}%
\bibitem [{\citenamefont {Sahu}\ \emph {et~al.}(2019)\citenamefont {Sahu},
  \citenamefont {Xu},\ and\ \citenamefont {Swingle}}]{sahu2019scrambling}%
  \BibitemOpen
  \bibfield  {author} {\bibinfo {author} {\bibfnamefont {S.}~\bibnamefont
  {Sahu}}, \bibinfo {author} {\bibfnamefont {S.}~\bibnamefont {Xu}},\ and\
  \bibinfo {author} {\bibfnamefont {B.}~\bibnamefont {Swingle}},\ }\bibfield
  {title} {\bibinfo {title} {Scrambling dynamics across a
  thermalization-localization quantum phase transition},\ }\href@noop {}
  {\bibfield  {journal} {\bibinfo  {journal} {Physical Review Letters}\
  }\textbf {\bibinfo {volume} {123}},\ \bibinfo {pages} {165902} (\bibinfo
  {year} {2019})}\BibitemShut {NoStop}%
\bibitem [{\citenamefont {Schuster}\ \emph {et~al.}(2021)\citenamefont
  {Schuster}, \citenamefont {Kobrin}, \citenamefont {Gao}, \citenamefont
  {Cong}, \citenamefont {Khabiboulline}, \citenamefont {Linke}, \citenamefont
  {Lukin}, \citenamefont {Monroe}, \citenamefont {Yoshida},\ and\ \citenamefont
  {Yao}}]{schuster2021many}%
  \BibitemOpen
  \bibfield  {author} {\bibinfo {author} {\bibfnamefont {T.}~\bibnamefont
  {Schuster}}, \bibinfo {author} {\bibfnamefont {B.}~\bibnamefont {Kobrin}},
  \bibinfo {author} {\bibfnamefont {P.}~\bibnamefont {Gao}}, \bibinfo {author}
  {\bibfnamefont {I.}~\bibnamefont {Cong}}, \bibinfo {author} {\bibfnamefont
  {E.~T.}\ \bibnamefont {Khabiboulline}}, \bibinfo {author} {\bibfnamefont
  {N.~M.}\ \bibnamefont {Linke}}, \bibinfo {author} {\bibfnamefont {M.~D.}\
  \bibnamefont {Lukin}}, \bibinfo {author} {\bibfnamefont {C.}~\bibnamefont
  {Monroe}}, \bibinfo {author} {\bibfnamefont {B.}~\bibnamefont {Yoshida}},\
  and\ \bibinfo {author} {\bibfnamefont {N.~Y.}\ \bibnamefont {Yao}},\
  }\bibfield  {title} {\bibinfo {title} {Many-body quantum teleportation via
  operator spreading in the traversable wormhole protocol},\ }\href@noop {}
  {\bibfield  {journal} {\bibinfo  {journal} {arXiv preprint arXiv:2102.00010}\
  } (\bibinfo {year} {2021})}\BibitemShut {NoStop}%
\bibitem [{\citenamefont {Li}\ \emph {et~al.}(2017)\citenamefont {Li},
  \citenamefont {Fan}, \citenamefont {Wang}, \citenamefont {Ye}, \citenamefont
  {Zeng}, \citenamefont {Zhai}, \citenamefont {Peng},\ and\ \citenamefont
  {Du}}]{li2017measuring}%
  \BibitemOpen
  \bibfield  {author} {\bibinfo {author} {\bibfnamefont {J.}~\bibnamefont
  {Li}}, \bibinfo {author} {\bibfnamefont {R.}~\bibnamefont {Fan}}, \bibinfo
  {author} {\bibfnamefont {H.}~\bibnamefont {Wang}}, \bibinfo {author}
  {\bibfnamefont {B.}~\bibnamefont {Ye}}, \bibinfo {author} {\bibfnamefont
  {B.}~\bibnamefont {Zeng}}, \bibinfo {author} {\bibfnamefont {H.}~\bibnamefont
  {Zhai}}, \bibinfo {author} {\bibfnamefont {X.}~\bibnamefont {Peng}},\ and\
  \bibinfo {author} {\bibfnamefont {J.}~\bibnamefont {Du}},\ }\bibfield
  {title} {\bibinfo {title} {Measuring out-of-time-order correlators on a
  nuclear magnetic resonance quantum simulator},\ }\href@noop {} {\bibfield
  {journal} {\bibinfo  {journal} {Physical Review X}\ }\textbf {\bibinfo
  {volume} {7}},\ \bibinfo {pages} {031011} (\bibinfo {year}
  {2017})}\BibitemShut {NoStop}%
\bibitem [{\citenamefont {Landsman}\ \emph {et~al.}(2019)\citenamefont
  {Landsman}, \citenamefont {Figgatt}, \citenamefont {Schuster}, \citenamefont
  {Linke}, \citenamefont {Yoshida}, \citenamefont {Yao},\ and\ \citenamefont
  {Monroe}}]{landsman2019verified}%
  \BibitemOpen
  \bibfield  {author} {\bibinfo {author} {\bibfnamefont {K.~A.}\ \bibnamefont
  {Landsman}}, \bibinfo {author} {\bibfnamefont {C.}~\bibnamefont {Figgatt}},
  \bibinfo {author} {\bibfnamefont {T.}~\bibnamefont {Schuster}}, \bibinfo
  {author} {\bibfnamefont {N.~M.}\ \bibnamefont {Linke}}, \bibinfo {author}
  {\bibfnamefont {B.}~\bibnamefont {Yoshida}}, \bibinfo {author} {\bibfnamefont
  {N.~Y.}\ \bibnamefont {Yao}},\ and\ \bibinfo {author} {\bibfnamefont
  {C.}~\bibnamefont {Monroe}},\ }\bibfield  {title} {\bibinfo {title} {Verified
  quantum information scrambling},\ }\href@noop {} {\bibfield  {journal}
  {\bibinfo  {journal} {Nature}\ }\textbf {\bibinfo {volume} {567}},\ \bibinfo
  {pages} {61} (\bibinfo {year} {2019})}\BibitemShut {NoStop}%
\bibitem [{\citenamefont {Blok}\ \emph {et~al.}(2021)\citenamefont {Blok},
  \citenamefont {Ramasesh}, \citenamefont {Schuster}, \citenamefont
  {O’Brien}, \citenamefont {Kreikebaum}, \citenamefont {Dahlen},
  \citenamefont {Morvan}, \citenamefont {Yoshida}, \citenamefont {Yao},\ and\
  \citenamefont {Siddiqi}}]{blok2021quantum}%
  \BibitemOpen
  \bibfield  {author} {\bibinfo {author} {\bibfnamefont {M.~S.}\ \bibnamefont
  {Blok}}, \bibinfo {author} {\bibfnamefont {V.~V.}\ \bibnamefont {Ramasesh}},
  \bibinfo {author} {\bibfnamefont {T.}~\bibnamefont {Schuster}}, \bibinfo
  {author} {\bibfnamefont {K.}~\bibnamefont {O’Brien}}, \bibinfo {author}
  {\bibfnamefont {J.-M.}\ \bibnamefont {Kreikebaum}}, \bibinfo {author}
  {\bibfnamefont {D.}~\bibnamefont {Dahlen}}, \bibinfo {author} {\bibfnamefont
  {A.}~\bibnamefont {Morvan}}, \bibinfo {author} {\bibfnamefont
  {B.}~\bibnamefont {Yoshida}}, \bibinfo {author} {\bibfnamefont {N.~Y.}\
  \bibnamefont {Yao}},\ and\ \bibinfo {author} {\bibfnamefont {I.}~\bibnamefont
  {Siddiqi}},\ }\bibfield  {title} {\bibinfo {title} {Quantum information
  scrambling on a superconducting qutrit processor},\ }\href@noop {} {\bibfield
   {journal} {\bibinfo  {journal} {Physical Review X}\ }\textbf {\bibinfo
  {volume} {11}},\ \bibinfo {pages} {021010} (\bibinfo {year}
  {2021})}\BibitemShut {NoStop}%
\bibitem [{\citenamefont {Wei}\ \emph {et~al.}(2019)\citenamefont {Wei},
  \citenamefont {Peng}, \citenamefont {Shtanko}, \citenamefont {Marvian},
  \citenamefont {Lloyd}, \citenamefont {Ramanathan}, \citenamefont {Cappellaro}
  \emph {et~al.}}]{wei2019emergent}%
  \BibitemOpen
  \bibfield  {author} {\bibinfo {author} {\bibfnamefont {K.~X.}\ \bibnamefont
  {Wei}}, \bibinfo {author} {\bibfnamefont {P.}~\bibnamefont {Peng}}, \bibinfo
  {author} {\bibfnamefont {O.}~\bibnamefont {Shtanko}}, \bibinfo {author}
  {\bibfnamefont {I.}~\bibnamefont {Marvian}}, \bibinfo {author} {\bibfnamefont
  {S.}~\bibnamefont {Lloyd}}, \bibinfo {author} {\bibfnamefont
  {C.}~\bibnamefont {Ramanathan}}, \bibinfo {author} {\bibfnamefont
  {P.}~\bibnamefont {Cappellaro}}, \emph {et~al.},\ }\bibfield  {title}
  {\bibinfo {title} {Emergent prethermalization signatures in out-of-time
  ordered correlations},\ }\href@noop {} {\bibfield  {journal} {\bibinfo
  {journal} {Physical Review Letters}\ }\textbf {\bibinfo {volume} {123}},\
  \bibinfo {pages} {090605} (\bibinfo {year} {2019})}\BibitemShut {NoStop}%
\bibitem [{\citenamefont {Martin}\ \emph {et~al.}(2022)\citenamefont {Martin},
  \citenamefont {Zhou}, \citenamefont {Leitao}, \citenamefont {Maskara},
  \citenamefont {Makarova}, \citenamefont {Gao}, \citenamefont {Zhu},
  \citenamefont {Park}, \citenamefont {Tyler}, \citenamefont {Park},
  \citenamefont {Choi},\ and\ \citenamefont {Lukin}}]{Martin22}%
  \BibitemOpen
  \bibfield  {author} {\bibinfo {author} {\bibfnamefont {L.~S.}\ \bibnamefont
  {Martin}}, \bibinfo {author} {\bibfnamefont {H.}~\bibnamefont {Zhou}},
  \bibinfo {author} {\bibfnamefont {N.~T.}\ \bibnamefont {Leitao}}, \bibinfo
  {author} {\bibfnamefont {N.}~\bibnamefont {Maskara}}, \bibinfo {author}
  {\bibfnamefont {O.}~\bibnamefont {Makarova}}, \bibinfo {author}
  {\bibfnamefont {H.}~\bibnamefont {Gao}}, \bibinfo {author} {\bibfnamefont
  {Q.-Z.}\ \bibnamefont {Zhu}}, \bibinfo {author} {\bibfnamefont
  {M.}~\bibnamefont {Park}}, \bibinfo {author} {\bibfnamefont {M.}~\bibnamefont
  {Tyler}}, \bibinfo {author} {\bibfnamefont {H.}~\bibnamefont {Park}},
  \bibinfo {author} {\bibfnamefont {S.}~\bibnamefont {Choi}},\ and\ \bibinfo
  {author} {\bibfnamefont {M.~D.}\ \bibnamefont {Lukin}},\ }\href@noop {}
  {\bibinfo {title} {Controlling local thermalization dynamics in a
  floquet-engineered dipolar ensemble}} (\bibinfo {year} {2022}),\ \Eprint
  {https://arxiv.org/abs/2209.09297} {arXiv:2209.09297 [quant-ph]} \BibitemShut
  {NoStop}%
\bibitem [{\citenamefont {Cappellaro}\ \emph {et~al.}(2007)\citenamefont
  {Cappellaro}, \citenamefont {Ramanathan},\ and\ \citenamefont
  {Cory}}]{cappellaro2007simulations}%
  \BibitemOpen
  \bibfield  {author} {\bibinfo {author} {\bibfnamefont {P.}~\bibnamefont
  {Cappellaro}}, \bibinfo {author} {\bibfnamefont {C.}~\bibnamefont
  {Ramanathan}},\ and\ \bibinfo {author} {\bibfnamefont {D.~G.}\ \bibnamefont
  {Cory}},\ }\bibfield  {title} {\bibinfo {title} {Simulations of information
  transport in spin chains},\ }\href@noop {} {\bibfield  {journal} {\bibinfo
  {journal} {Physical review letters}\ }\textbf {\bibinfo {volume} {99}},\
  \bibinfo {pages} {250506} (\bibinfo {year} {2007})}\BibitemShut {NoStop}%
\bibitem [{\citenamefont {Cappellaro}\ \emph {et~al.}(2011)\citenamefont
  {Cappellaro}, \citenamefont {Viola},\ and\ \citenamefont
  {Ramanathan}}]{cappellaro2011coherent}%
  \BibitemOpen
  \bibfield  {author} {\bibinfo {author} {\bibfnamefont {P.}~\bibnamefont
  {Cappellaro}}, \bibinfo {author} {\bibfnamefont {L.}~\bibnamefont {Viola}},\
  and\ \bibinfo {author} {\bibfnamefont {C.}~\bibnamefont {Ramanathan}},\
  }\bibfield  {title} {\bibinfo {title} {Coherent-state transfer via highly
  mixed quantum spin chains},\ }\href@noop {} {\bibfield  {journal} {\bibinfo
  {journal} {Physical Review A}\ }\textbf {\bibinfo {volume} {83}},\ \bibinfo
  {pages} {032304} (\bibinfo {year} {2011})}\BibitemShut {NoStop}%
\bibitem [{\citenamefont {Ramanathan}\ \emph {et~al.}(2011)\citenamefont
  {Ramanathan}, \citenamefont {Cappellaro}, \citenamefont {Viola},\ and\
  \citenamefont {Cory}}]{ramanathan2011experimental}%
  \BibitemOpen
  \bibfield  {author} {\bibinfo {author} {\bibfnamefont {C.}~\bibnamefont
  {Ramanathan}}, \bibinfo {author} {\bibfnamefont {P.}~\bibnamefont
  {Cappellaro}}, \bibinfo {author} {\bibfnamefont {L.}~\bibnamefont {Viola}},\
  and\ \bibinfo {author} {\bibfnamefont {D.~G.}\ \bibnamefont {Cory}},\
  }\bibfield  {title} {\bibinfo {title} {Experimental characterization of
  coherent magnetization transport in a one-dimensional spin system},\
  }\href@noop {} {\bibfield  {journal} {\bibinfo  {journal} {New Journal of
  Physics}\ }\textbf {\bibinfo {volume} {13}},\ \bibinfo {pages} {103015}
  (\bibinfo {year} {2011})}\BibitemShut {NoStop}%
\bibitem [{\citenamefont {Rufeil-Fiori}\ \emph {et~al.}(2009)\citenamefont
  {Rufeil-Fiori}, \citenamefont {S{\'a}nchez}, \citenamefont {Oliva},
  \citenamefont {Pastawski},\ and\ \citenamefont
  {Levstein}}]{rufeil2009effective}%
  \BibitemOpen
  \bibfield  {author} {\bibinfo {author} {\bibfnamefont {E.}~\bibnamefont
  {Rufeil-Fiori}}, \bibinfo {author} {\bibfnamefont {C.~M.}\ \bibnamefont
  {S{\'a}nchez}}, \bibinfo {author} {\bibfnamefont {F.~Y.}\ \bibnamefont
  {Oliva}}, \bibinfo {author} {\bibfnamefont {H.~M.}\ \bibnamefont
  {Pastawski}},\ and\ \bibinfo {author} {\bibfnamefont {P.~R.}\ \bibnamefont
  {Levstein}},\ }\bibfield  {title} {\bibinfo {title} {Effective one-body
  dynamics in multiple-quantum nmr experiments},\ }\href@noop {} {\bibfield
  {journal} {\bibinfo  {journal} {Physical Review A}\ }\textbf {\bibinfo
  {volume} {79}},\ \bibinfo {pages} {032324} (\bibinfo {year}
  {2009})}\BibitemShut {NoStop}%
\bibitem [{\citenamefont {Zhang}\ \emph {et~al.}(2009)\citenamefont {Zhang},
  \citenamefont {Cappellaro}, \citenamefont {Antler}, \citenamefont {Pepper},
  \citenamefont {Cory}, \citenamefont {Dobrovitski}, \citenamefont
  {Ramanathan},\ and\ \citenamefont {Viola}}]{zhang2009nmr}%
  \BibitemOpen
  \bibfield  {author} {\bibinfo {author} {\bibfnamefont {W.}~\bibnamefont
  {Zhang}}, \bibinfo {author} {\bibfnamefont {P.}~\bibnamefont {Cappellaro}},
  \bibinfo {author} {\bibfnamefont {N.}~\bibnamefont {Antler}}, \bibinfo
  {author} {\bibfnamefont {B.}~\bibnamefont {Pepper}}, \bibinfo {author}
  {\bibfnamefont {D.~G.}\ \bibnamefont {Cory}}, \bibinfo {author}
  {\bibfnamefont {V.~V.}\ \bibnamefont {Dobrovitski}}, \bibinfo {author}
  {\bibfnamefont {C.}~\bibnamefont {Ramanathan}},\ and\ \bibinfo {author}
  {\bibfnamefont {L.}~\bibnamefont {Viola}},\ }\bibfield  {title} {\bibinfo
  {title} {Nmr multiple quantum coherences in quasi-one-dimensional spin
  systems: Comparison with ideal spin-chain dynamics},\ }\href@noop {}
  {\bibfield  {journal} {\bibinfo  {journal} {Physical Review A}\ }\textbf
  {\bibinfo {volume} {80}},\ \bibinfo {pages} {052323} (\bibinfo {year}
  {2009})}\BibitemShut {NoStop}%
\bibitem [{\citenamefont {Yen}\ and\ \citenamefont
  {Pines}(1983{\natexlab{a}})}]{Yen83}%
  \BibitemOpen
  \bibfield  {author} {\bibinfo {author} {\bibfnamefont {Y.-S.}\ \bibnamefont
  {Yen}}\ and\ \bibinfo {author} {\bibfnamefont {A.}~\bibnamefont {Pines}},\
  }\bibfield  {title} {\bibinfo {title} {Multiple-quantum nmr in solids},\
  }\href {https://doi.org/10.1063/1.445185} {\bibfield  {journal} {\bibinfo
  {journal} {J. Comp. Phys.}\ }\textbf {\bibinfo {volume} {78}},\ \bibinfo
  {pages} {3579} (\bibinfo {year} {1983}{\natexlab{a}})}\BibitemShut {NoStop}%
\bibitem [{\citenamefont {Kaur}\ and\ \citenamefont
  {Cappellaro}(2012)}]{kaur2012initialization}%
  \BibitemOpen
  \bibfield  {author} {\bibinfo {author} {\bibfnamefont {G.}~\bibnamefont
  {Kaur}}\ and\ \bibinfo {author} {\bibfnamefont {P.}~\bibnamefont
  {Cappellaro}},\ }\bibfield  {title} {\bibinfo {title} {Initialization and
  readout of spin chains for quantum information transport},\ }\href@noop {}
  {\bibfield  {journal} {\bibinfo  {journal} {New Journal of Physics}\ }\textbf
  {\bibinfo {volume} {14}},\ \bibinfo {pages} {083005} (\bibinfo {year}
  {2012})}\BibitemShut {NoStop}%
\bibitem [{\citenamefont {Yen}\ and\ \citenamefont
  {Pines}(1983{\natexlab{b}})}]{yen1983multiple}%
  \BibitemOpen
  \bibfield  {author} {\bibinfo {author} {\bibfnamefont {Y.-S.}\ \bibnamefont
  {Yen}}\ and\ \bibinfo {author} {\bibfnamefont {A.}~\bibnamefont {Pines}},\
  }\bibfield  {title} {\bibinfo {title} {Multiple-quantum nmr in solids},\
  }\href@noop {} {\bibfield  {journal} {\bibinfo  {journal} {The Journal of
  chemical physics}\ }\textbf {\bibinfo {volume} {78}},\ \bibinfo {pages}
  {3579} (\bibinfo {year} {1983}{\natexlab{b}})}\BibitemShut {NoStop}%
\bibitem [{\citenamefont {Ajoy}\ and\ \citenamefont
  {Cappellaro}(2013)}]{ajoy2013quantum}%
  \BibitemOpen
  \bibfield  {author} {\bibinfo {author} {\bibfnamefont {A.}~\bibnamefont
  {Ajoy}}\ and\ \bibinfo {author} {\bibfnamefont {P.}~\bibnamefont
  {Cappellaro}},\ }\bibfield  {title} {\bibinfo {title} {Quantum simulation via
  filtered hamiltonian engineering: Application to perfect quantum transport in
  spin networks},\ }\href@noop {} {\bibfield  {journal} {\bibinfo  {journal}
  {Physical review letters}\ }\textbf {\bibinfo {volume} {110}},\ \bibinfo
  {pages} {220503} (\bibinfo {year} {2013})}\BibitemShut {NoStop}%
\bibitem [{\citenamefont {Comodi}\ \emph {et~al.}(2001)\citenamefont {Comodi},
  \citenamefont {Liu}, \citenamefont {Zanazzi},\ and\ \citenamefont
  {Montagnoli}}]{Comodi01}%
  \BibitemOpen
  \bibfield  {author} {\bibinfo {author} {\bibfnamefont {P.}~\bibnamefont
  {Comodi}}, \bibinfo {author} {\bibfnamefont {Y.}~\bibnamefont {Liu}},
  \bibinfo {author} {\bibfnamefont {P.}~\bibnamefont {Zanazzi}},\ and\ \bibinfo
  {author} {\bibfnamefont {M.}~\bibnamefont {Montagnoli}},\ }\bibfield  {title}
  {\bibinfo {title} {Structural and vibrational behaviour of fluorapatite with
  pressure. part i: in situ single-crystal x-ray diffraction investigation},\
  }\href {https://link.springer.com/article/10.1007/s002690100154} {\bibfield
  {journal} {\bibinfo  {journal} {Physics and Chemistry of Minerals}\ }\textbf
  {\bibinfo {volume} {28}},\ \bibinfo {pages} {219} (\bibinfo {year}
  {2001})}\BibitemShut {NoStop}%
\bibitem [{\citenamefont {Deutsch}(1991)}]{deutsch1991quantum}%
  \BibitemOpen
  \bibfield  {author} {\bibinfo {author} {\bibfnamefont {J.~M.}\ \bibnamefont
  {Deutsch}},\ }\bibfield  {title} {\bibinfo {title} {Quantum statistical
  mechanics in a closed system},\ }\href@noop {} {\bibfield  {journal}
  {\bibinfo  {journal} {Physical review a}\ }\textbf {\bibinfo {volume} {43}},\
  \bibinfo {pages} {2046} (\bibinfo {year} {1991})}\BibitemShut {NoStop}%
\bibitem [{\citenamefont {Rigol}\ \emph {et~al.}(2008)\citenamefont {Rigol},
  \citenamefont {Dunjko},\ and\ \citenamefont
  {Olshanii}}]{rigol2008thermalization}%
  \BibitemOpen
  \bibfield  {author} {\bibinfo {author} {\bibfnamefont {M.}~\bibnamefont
  {Rigol}}, \bibinfo {author} {\bibfnamefont {V.}~\bibnamefont {Dunjko}},\ and\
  \bibinfo {author} {\bibfnamefont {M.}~\bibnamefont {Olshanii}},\ }\bibfield
  {title} {\bibinfo {title} {Thermalization and its mechanism for generic
  isolated quantum systems},\ }\href@noop {} {\bibfield  {journal} {\bibinfo
  {journal} {Nature}\ }\textbf {\bibinfo {volume} {452}},\ \bibinfo {pages}
  {854} (\bibinfo {year} {2008})}\BibitemShut {NoStop}%
\bibitem [{Note2()}]{Note2}%
  \BibitemOpen
  \bibinfo {note} {The $\pi $ pulse is implemented by changing the last $\pi
  $/2 pulse of WAHUHA8 from -y to y. Compared to physically applying a $\pi $
  pulse, the phase change does not elongate the sequence thus is more
  robust.}\BibitemShut {Stop}%
\bibitem [{\citenamefont {Suter}\ \emph {et~al.}(1987)\citenamefont {Suter},
  \citenamefont {Liu}, \citenamefont {Baum},\ and\ \citenamefont
  {Pines}}]{Suter87}%
  \BibitemOpen
  \bibfield  {author} {\bibinfo {author} {\bibfnamefont {D.}~\bibnamefont
  {Suter}}, \bibinfo {author} {\bibfnamefont {S.}~\bibnamefont {Liu}}, \bibinfo
  {author} {\bibfnamefont {J.}~\bibnamefont {Baum}},\ and\ \bibinfo {author}
  {\bibfnamefont {A.}~\bibnamefont {Pines}},\ }\bibfield  {title} {\bibinfo
  {title} {Multiple quantum nmr excitation with a one-quantum hamiltonian},\
  }\href {https://doi.org/10.1016/0301-0104(87)80023-X} {\bibfield  {journal}
  {\bibinfo  {journal} {Chem. Phys.}\ }\textbf {\bibinfo {volume} {114}},\
  \bibinfo {pages} {103 } (\bibinfo {year} {1987})}\BibitemShut {NoStop}%
\bibitem [{\citenamefont {Cho}\ \emph {et~al.}(2003)\citenamefont {Cho},
  \citenamefont {Cory},\ and\ \citenamefont {Ramanathan}}]{Cho03}%
  \BibitemOpen
  \bibfield  {author} {\bibinfo {author} {\bibfnamefont {H.}~\bibnamefont
  {Cho}}, \bibinfo {author} {\bibfnamefont {D.~G.}\ \bibnamefont {Cory}},\ and\
  \bibinfo {author} {\bibfnamefont {C.}~\bibnamefont {Ramanathan}},\ }\bibfield
   {title} {\bibinfo {title} {Spin counting experiments in the dipolar-ordered
  state},\ }\href {https://doi.org/10.1063/1.1538244} {\bibfield  {journal}
  {\bibinfo  {journal} {J. Comp. Phys.}\ }\textbf {\bibinfo {volume} {118}},\
  \bibinfo {pages} {3686} (\bibinfo {year} {2003})}\BibitemShut {NoStop}%
\bibitem [{\citenamefont {Wigner}(2012)}]{Wigner12}%
  \BibitemOpen
  \bibfield  {author} {\bibinfo {author} {\bibfnamefont {E.}~\bibnamefont
  {Wigner}},\ }\href@noop {} {\emph {\bibinfo {title} {Group theory: and its
  application to the quantum mechanics of atomic spectra}}},\ Vol.~\bibinfo
  {volume} {5}\ (\bibinfo  {publisher} {Elsevier},\ \bibinfo {year}
  {2012})\BibitemShut {NoStop}%
\bibitem [{\citenamefont {van Beek}\ \emph {et~al.}(2005)\citenamefont {van
  Beek}, \citenamefont {Carravetta}, \citenamefont {Antonioli},\ and\
  \citenamefont {Levitt}}]{vanBeek05}%
  \BibitemOpen
  \bibfield  {author} {\bibinfo {author} {\bibfnamefont {J.~D.}\ \bibnamefont
  {van Beek}}, \bibinfo {author} {\bibfnamefont {M.}~\bibnamefont
  {Carravetta}}, \bibinfo {author} {\bibfnamefont {G.~C.}\ \bibnamefont
  {Antonioli}},\ and\ \bibinfo {author} {\bibfnamefont {M.~H.}\ \bibnamefont
  {Levitt}},\ }\bibfield  {title} {\bibinfo {title} {Spherical tensor analysis
  of nuclear magnetic resonance signals},\ }\href
  {https://doi.org/10.1063/1.1943947} {\bibfield  {journal} {\bibinfo
  {journal} {J. Comp. Phys.}\ }\textbf {\bibinfo {volume} {122}},\ \bibinfo
  {eid} {244510} (\bibinfo {year} {2005})}\BibitemShut {NoStop}%
\end{thebibliography}%

%
\end{document}